\documentclass{aa}
\usepackage{epsfig}
\usepackage[varg]{txfonts}
\usepackage{lscape}
\usepackage{multirow}
\usepackage{multicol}
\usepackage{subfigure}
\usepackage{color}

\def\aap{A\&A }

\def\apjs{ApJS }

\def\apjl{ApJL }

\def\apjs{ApJS }

\newcommand{\eg}{{\it e.g. }}
\newcommand{\ie}{{\it i.e. }}

\begin{document}

\title{The truncated and evolving inner accretion disc of the black hole GX 339$-$4}
\author{D.~S. Plant\inst{\ref{inst1}}\and R.~P. Fender\inst{\ref{inst2},\ref{inst1}}\and G. Ponti\inst{\ref{inst3}}\and T. Mu\~{n}oz-Darias\inst{\ref{inst2},\ref{inst1}}\and M. Coriat\inst{\ref{inst1}}}

\institute{School of Physics and Astronomy, University of Southampton, Highfield, Southampton, SO17 1BJ, United Kingdom\label{inst1}
\thanks{\emph{Present address:} Department of Physics, Astrophysics, University of Oxford, Keble Road, Oxford, OX1 3RH, United Kingdom}
\\ \email{Daniel.Plant@astro.ox.ac.uk}
\and
Department of Physics, Astrophysics, University of Oxford, Keble Road, Oxford, OX1 3RH, United Kingdom\label{inst2}
\and
Max Planck Institute fur Extraterrestriche Physik, 85748, Garching, Germany\label{inst3}
}

\date{Received ... / Accepted ...}

\abstract{The nature of accretion onto stellar mass black holes in the low/hard state remains unresolved, with some evidence suggesting that the inner accretion disc is truncated and replaced by a hot flow. However, the detection of relativistic broadened Fe emission lines, even at relatively low luminosities, seems to require an accretion disc extending fully to its innermost stable circular orbit. Modelling such features is however highly susceptible to degeneracies, which could easily bias any interpretation. We present the first systematic study of the Fe line region to track how the inner accretion disc evolves in the low/hard state of the black hole GX 339$-$4. Our four observations display increased broadening of the Fe line over two magnitudes in luminosity, which we use to track any variation of the disc inner radius. We find that the disc extends closer to the black hole at higher luminosities, but is consistent with being truncated throughout the entire low/hard state, a result which renders black hole spin estimates inaccurate at these stages of the outburst. Furthermore, we show that the evolution of our spectral inner disc radius estimates corresponds very closely to the trend of the break frequency in Fourier power spectra, supporting the interpretation of a truncated and evolving disc in the hard state.}

\maketitle


\section{Introduction}\label{intro}
Black Hole X-ray Binaries (BHXRBs) are, in the main, transient, spending the majority of their lifetime in a `quiescent' state punctuated by intense outbursts, which can last for months or even years. These outbursts are driven by accretion disc instabilities (see \citealt{Meyer81, Coriat12} and references therein) and we observe distinct spectral states revealed through the relative strength of their respective soft and hard X-ray emission \citep{Remillard06, Done07, Belloni11, Fender12}.

An outburst commences in the low/hard (hereafter `hard') state, characterised by its dominant power-law component and high level of aperiodic variability. The power-law is believed to originate as a result of inverse Compton scattering of `seed' photons in a thermalised, optically thin, `corona', with a temperature of $\sim100$\,keV. Present, albeit weakly, is a soft excess, generally attributed to arise from the cool and dim accretion disc. This is starkly juxtaposed with the high/soft (hereafter `soft') state where a quasi-blackbody component (kT $\sim1$\,keV) is now dominant and the hard power-law has steepened, contributing very little to the overall luminosity.

These two distinct states are ubiquitously associated with two different types of outflow: winds and jets. Narrow absorption features have been detected in numerous systems and high-resolution spectroscopy has attributed them to an accretion disc wind \citep{Ueda98, Lee02, Miller06c, DT07}. Whilst winds are thought to be ubiquitous in BHXRBs, they are only observed in the soft state (\citealt{Ponti12}; see \citealt{DT12} for a recent review) whereas a similar, but reversed, pattern is observed for the jet, which is quenched in the soft state \citep{Russell11} but steady in the hard state \citep{Fender04}. These two outflow phenomena may be physically linked \citep{Neilsen09} or just symptoms of the state change. The transition between the hard and soft states is often referred to as the `intermediate' phase, again split into two distinct regimes: the hard-intermediate (HIMS) and soft-intermediate (SIMS) states. In the HIMS the spectrum has softened as a result of a steeper power-law index (now as high as $2.5$, whereas $\sim$$1.6$ in the hard state) and an increased thermal disc component. Furthermore, it displays variability similar to the hard state, with increased characteristic frequencies \citep{VDK06,Belloni11}. The SIMS spectrum is slightly softer, but not significantly; however, the timing properties change abruptly displaying variability as little as a few \%, marking a clearly different state.

Also present and superimposed upon the continuum are reflection features emerging from irradiation of the accretion disc by the up-scattered seed photons \citep{Reynolds03, Miller07,Fabian10}. Emission from the inner regions of the accretion flow will undergo both special and general relativistic effects leading to a broadened and skewed profile \citep{Fabian89}. Such a profile is a result of the strong dependance of relativistic effects with distance from the central Black Hole (BH), allowing it to be used as a diagnostic of the inner accretion disc, specifically the inner radius. Furthermore, assuming the disc is at the innermost stable circular orbit (ISCO), one can essentially use the inner radius to estimate the spin of the BH, since the ISCO evolves from $6\,r_{\rm g}$ to $\sim$$1.2\,r_{\rm g}$ for the full range of prograde spin \citep{Bardeen72, Thorne74}. However, even given a fully robust model, there is still one rather large assumption here - that the inner disc has extended all the way to the ISCO.

It is widely accepted that the inner accretion disc extends fully to the ISCO for a large fraction of the soft state \citep{Gierlinski04, Steiner10}. In quiescence, however, the disc is predicted and found to recede as the accretion rate diminishes \citep{McClintock95, Narayan95, Narayan96, Esin97, McClintock01, Esin01, McClintock03}. At some stage, therefore, the inner accretion disc must evolve and extend itself closer to the BH, a topic which has become strongly debated in recent times.

A truncated inner accretion disc provides a geometric interpretation to explain the spectral changes observed in BHXRBs \citep{Done07,Plant14}. Here, the standard optically thick and geometrically thin accretion disc is replaced in the inner regions by what is essentially its converse; a hot, optically thin, geometrically thick flow \citep{Esin97}. As the accretion rate increases, the truncation radius of the disc is expected to decrease, eventually extending fully to the ISCO. However, there are two distinct claims in the literature strongly arguing against this, even at relatively low luminosities in the hard state. (1) The detection of significantly broadened Fe lines at low accretion rates \citep{Miller06b, Reis08, Reis10} offer evidence of relativistic effects at or near to the ISCO. However, a number of works have re-analysed these observations and instead found evidence for inner disc truncation \citep{Done06, Yamada09,Done10,Kolehmainen13}. Furthermore, \cite{Tomsick09}, \cite{Shidatsu11} and \cite{Petrucci14} have found further evidence for disc truncation in additional hard state observations with \emph{Suzaku}. (2) A geometrically thin disc with a fixed inner radius will closely follow the $L\propto T^4$ relation \citep{Gierlinski04, Dunn11}, and hence deviations from this can reveal truncation of the inner disc. The debate remains unresolved with arguments for \citep{Gierlinski08, Tomsick08, Cabanac09} and against \citep{Rykoff07, Miller06, Reis10} disc truncation. Furthermore, there is evidence that the disc may also be truncated in the hard intermediate state \citep{Kubota04, Done06b, Tamura12}, or at the ISCO \citep{Hiemstra11}.

GX 339-4 is a key source in the inner disc truncation debate. Investigations by \cite{Miller06b} and \cite{Reis08} of strongly broadened Fe lines have suggested that the disc is at the ISCO in the hard state. However, \cite{Done10} demonstrated that these data can be largely affected by photon pile-up, and revealed that the emission line could be significantly narrower. Studies by \cite{Tomsick09}, \cite{Shidatsu11} and \cite{Petrucci14} have examined the Fe line region further at lower luminosities, providing strong evidence for a truncated disc, and hints at a correlation between the inner radius estimates and luminosity. Currently, therefore, our understanding of the state of the inner accretion disc in the hard state of GX 339-4 is uncertain. It is essential that we are confident of how the inner disc evolves in the hard state in order for us to measure the BH spin through the X-ray reflection method. This requires that the inner disc is at the ISCO and relies heavily on studies of the hard state, where irradiation of the disc is high and the underlying continuum around $6$\,keV is relatively simple (see \eg \citealt{Kolehmainen11} for issues with softer spectral states).

All of the observations used in this study have previously been analysed in other works, but directly comparing these results can be misleading. Firstly, the parameters derived from reflection modelling can be highly degenerate and can significantly affect conclusions if not accounted for. This can be observed through the range of inclination angles fitted in the studies mentioned previously. The inclination angle of GX 339-4 has not been accurately measured, and freely fitting this parameter can crucially affect the inner radius estimate \citep{Tomsick09}. Additionally, assorted interpretations of the spectra have lead to a variety of models being applied which can again skew results. We therefore present the first systematic study of evolution of the Fe line region in the hard state of GX 339-4 by fitting our selected observations simultaneously, enabling potentially degenerate parameters to be tied. This allows us to keep our analysis as consistent as possible and significantly remove the issues described before by inspecting the relative change in the inner disc evolution, specifically the ionisation and inner radius parameters. Our focus in this investigation is only GX 339-4 since it is the best sampled transient BH at the resolution we require in the hard state (see \S\ref{obs_data}). We do note, however, that although transient BHs generally display the same behaviour in outburst, it may be the case that the inner disc does not always vary how we find in this study.

We begin by introducing our observation selection criteria and data reduction process in \S\ref{obs_data}. We then firstly analyse the continuum emission (\S\ref{continuum}) and then proceed to fitting the Fe line region, utilising both line and self-consistent reflection modelling techniques (\S\ref{line_model} - \ref{full_ref}). Our results indicate that even at relatively high luminosities the disc appears to still be recessed in the hard state, in addition to a strong trend of evolution towards the ISCO at higher luminosity. We then perform a detailed Markov Chain Monte Carlo analysis in \S\ref{Sec:MCMC} to confirm the significance of this result. Additionally, we analyse whether this trend is purely based on luminosity by comparing observations in the rise and decay of the hard state at similar flux levels (\S\ref{rise_decay}). Next in \S\ref{timing_sec} we investigate any correlation our results may have with the power spectra break frequency, which is believed to also track how the inner disc evolves. Finally, in \S\ref{full_test} we examine how our results may be effected by the spectral bandpass we employ, the fitted inclination of the system, and the chosen emissivity profile. We then summarise our findings and discuss potential caveats, plus implications for the truncated disc model and BH spin estimates, in \S\ref{discussion} and \S\ref{conclusions}.


\section{Observations and data reduction}\label{obs_data}
We utilise the Fe line region as a probe of disc evolution with luminosity. To undertake such a task, which requires high precision, the work presented here considers only high resolution observations with a large effective area in the Fe K band: \emph{XMM-Newton} and \emph{Suzaku}. Our focus is the hard and hard-intermediate phases of the outburst, since the state of the inner disc is unclear during this phase, and has subsequently lead to much debate in recent years. This is also ideal since a study by \cite{Kolehmainen11} displayed that in the case of the disc and power-law components having equal fluxes around the Fe K range, i.e. the soft and soft-intermediate states, the ability to determine the width of the Fe profile is severely hampered. Whilst the data we ignore restricts our sample to probe the inner disc evolution, the inner disc is already well regarded to have extended to the ISCO in the soft states \citep{Gierlinski04,Steiner10}. 

The remainder of this section introduces our data reduction procedure, and the five observations found in the archives which met our criteria are described in Table \ref{observations}.

\renewcommand{\arraystretch}{1.2}
\begin{table*}
\centering
\caption{Observation log for the X-ray datasets used in this study.}
\begin{tabular}{llllll}\hline\hline
					& ObsID			& Date			& State						& Net Count Rate (cts/s)			& Exposure					\\\hline	
1					& 403067010	& 2008-09-24    	& Hard (Decay)				& $2\pm 0.01$					& 105\,ks					\\
2					& 0605610201	& 2009-03-26 	& Hard (Rise)				& $132\pm 0.1$					& 32\,ks						\\
2b					& 405063010	& 2011-02-11 	& HIMS (Decay)				& $20\pm 0.02$ (29) 				& 22\,ks						\\
\multirow{2}{*}{3}		& 0204730201	& 2004-03-16 	& \multirow{2}{*}{Hard (Rise)}	& \multirow{2}{*}{$259\pm 0.1$}	& \multirow{2}{*}{155\,ks}		\\
					& 0204730301	& 2004-03-18 	& 							&								&							\\
4					& 0654130401	& 2010-03-28 	& Hard (Rise)				& $352\pm 0.2$ (988)				& 25\,ks						\\\hline
\end{tabular}
\tablefoot{Observations 2, 3, 4 were taken with \emph{XMM-Newton}, whilst 1 and 2b are \emph{Suzaku} datasets. Exposures correspond to that remaining after our reduction process, and hence used in our analysis. Values in brackets refer to the count rate before any pile-up removal was applied.}
\label{observations}
\end{table*}


\subsection{XMM-Newton}\label{xmm}
For the reduction of \emph{XMM-Newton} observations we employed the Science Analysis Software (SAS) version 11.0. In this study we do not consider data from the EPIC-MOS camera \citep{Turner01}, and instead restrict ourselves to the EPIC-pn \citep{Struder01} for which the three Observations were all taken in `timing' mode. In \S\ref{Sec:Pile-up} we outline our reasons for this and how the effects of pile-up were mitigated. We apply the standard tools to create response and ancillary files (\textsc{rmfgen} and \textsc{arfgen}) and use only single and double events (PATTERN$\leq$4), whilst ignoring bad pixels with \#XMMEA\_EP and FLAG==0. All spectra were extracted from a region in RAWX [31:45] and were binned using the FTOOL \textsc{grppha} to have at least $20$ counts per channel. Finally, we found all background regions were contaminated, indicated by spectra clearly following that of the source. However, due to the high source flux in all the observations we found it acceptable to proceed without implementing any background subtraction.

In \S\ref{Sec:Fe_fits}--\ref{rise_decay} we fit the EPIC-pn spectra in the 4--10\,keV band. This enables simultaneous fitting of the spectra with tied parameters and a well calibrated bandpass in a reasonable timescale. We then check the consistency of these results in \S\ref{full_test} by fitting each observation individually in the full bandpass of 1.3--10 keV, as is the standard method for X-ray studies of BHXRBs. Many ``soft excesses" have been reported in binaries with moderate to high column densities, the origin of which remains uncertain and appears to not be limited to the timing mode (see the XMM-Newton Calibration Technical Note (0083)$^1$ and references therein). We also find a significant residual below 1 keV which severely affects the fit, and hence we follow \cite{Hiemstra11} and \cite{Reis11} in setting a 1.3\,keV lower limit to the bandpass. We also ignore the 1.75--2.35\,keV region which contains strong features, likely to be of instrumental origin.

Charge-transfer inefficiency (CTI) results in a gain shift, and hence affects the energy spectrum, and can be corrected by the SAS task \textsc{epfast}. More importantly, if not allowed for CTI can shift the Fe K$\alpha$ profile (see Fig.\,22 in the XMM-Newton Calibration Technical Note (0083)\footnote{http://xmm.vilspa.esa.es/external/xmm\_sw\_cal/calib/\\documentation.shtml}). In a recent study, \cite{Walton12} showed that the rate-dependant \textsc{epfast} correction could also lead to an incorrect profile; however, they made use of a `burst' mode observation at a very high count rate where the correction may not be suitable. We follow the current recommendation and apply \textsc{epfast} to all our data, but we also note that the low count rate of our observations did not result in any noticeable changes. This has also been found at similar count rates in the EPIC-pn timing mode (see {\it e.g.} \citealt{Chiang12})

\subsubsection{Mitigating pile-up}\label{Sec:Pile-up}

Pile-up occurs when several photons hit two neighbouring (pattern pile-up) or the same (photon pile-up) pixel during one read-out cycle. If this happens the events are counted as one single event carrying an energy of the sum of the two or more incident photons, leading to a loss of flux or spectral hardening depending on whether the energy-rejection threshold is exceeded. It is, therefore, a very important issue when observing bright sources, such as BHXRBs. The effect of pile-up upon the Fe line profile is uncertain. Simulations by \cite{Miller10} suggest that it leads to the narrowing of line profiles, but spectral studies appear to display broadening \citep{Done10, Ng10}.

In this study we only utilise the EPIC-pn camera, and in the main this is to ensure pile-up is mitigated. For Observation 4 the EPIC-MOS data cannot be studied since only the EPIC-pn camera was operated. In Observation 3 the MOS was active and suffers significantly from pile-up. This dataset was taken in the full-frame imaging mode, for which WebPIMMS (using values from Table \ref{table_continuum}) predicts a count rate of over 100 times the nominal pile-up limit of 0.7\,cts/s recommended in the XMM Users Handbook. Since such a large annulus region is required to solve for this, the resultant spectra has less than 2\,\% of the counts registered in the EPIC-pn dataset between 4--10\,keV, which we later determine to be pile-up free. The MOS dataset for Observation 2 was taken in timing mode, which should be free of pile-up. However, given that the pn is certainly free of pile-up for this observation, and is a much superior instrument at 6\,keV, we decided the pn was the best option for our study. A further motivation for this is that using only the pn reduces the calibration dependance of our results.

The default check for pile-up is to examine the registered count rate and compare it to the nominal limit for pile-up recommended by the XMM team. For the EPIC-pn timing mode this is 800\,cts/s, of which Observations 2 and 3 are well below (see Table \ref{observations}), although Observation 4 exceeds this threshold. A more detailed test comes from the SAS task \textsc{epatplot}, which displays how the distribution of single and double-pixel (and higher) events compare to their expected values. If the two diverge this acts as strong evidence for pile-up, and hence we apply this to all three \emph{XMM-Newton} observations. Again, Observations 2 and 3 show no evidence for pile-up, although Observation 4 does. We find that the removal of the inner 3 RAWX columns in Observation 4 resolved the event distribution in \textsc{epatplot}, and also reduces the net count rate to a very acceptable 352\,cts/s.

In a final test we compared the spectra when different inner column regions are removed, whereby if pile-up is present one should see differences in the Fe line and spectral hardening. Observations 2 and 3 show little difference when inner columns are removed, with just some minor softening above 8 keV, which is likely to be due to calibration uncertainties in the determination of the effective area for PSF with central holes (this has been confirmed by the XMM-Newton EPIC calibration team). In contrast Observation 4 shows a significantly different spectra for a full strip compared to that with the three inner columns removed. Comparisons of three to five and seven columns being removed again only show some softening above 8\,keV, and thus confirm that the removal of three inner columns in Observation 4 is sufficient to mitigate pile-up. We therefore confirm that the three EPIC-pn spectra used in this study are free from the effects of pile-up.


\subsection{Suzaku}
Suzaku carries four X-ray Imaging Spectrometer detectors \cite[XIS;][]{Koyama07}, one of which is `back-illuminated' (BI) in addition to the three `front-illuminated' (FI) ones. Each covers the 0.2--12\,keV band. On 2006 November 9, XIS2, one of the FI detectors, failed and hence will not be considered in this work. Furthermore the Hard X-ray Telescope \citep[HXD;][]{Takahashi07, Kokubun07} extends coverage to the 10--70\,keV (PIN) and 50--600\,keV (GSO) regions respectively.

We processed the unfiltered event files following the \emph{Suzaku Data Reduction Guide} using the HEADAS v6.11.1 software package. We produced clean event files using the FTOOL \textsc{aepipeline}, applying the calibration products (HXD20110913, XIS20120209 and XRT20110630). Suzaku undergoes wobbling due to thermal flexing leading to a blurring of the image. We ran the script \textsc{aeattcor.sl}\footnote{http://space.mit.edu/CXC/software/suzaku/aeatt.html} to create a new attitude file, which was then applied to each clean event file using the FTOOL XISCOORD. For the XIS XSELECT was then used to extract the spectral and background products. A source region with a radius of $200$ pixels (1 pixel = 1.04 arcsec) centred on the image peak was used for all observations. Observation 2b employed the $1/4$ window mode, and therefore the extraction region is larger than the window itself, and hence the effective region extracted is an intersection of our circle with a rectangle of $1024\times256$ pixels. Background events were extracted using a circle of $100$ pixels located away from the source. Observation 2b is named such because it is at a similar flux level to Observation 2. The spectral hardness of Observation 2b is significantly softer (0.80 vs \textgreater\,0.93 for the hard state observations)\footnote{The spectral hardness was calculated using the ratio of the 6--10 and 3--6\,keV model flux.}, thus we identify it as being in the HIMS (see also \citealt{Petrucci14}).

We also check for the effects of photon pile-up using the tool \textsc{pile\_estimate.sl}\footnote{http://space.mit.edu/CXC/software/suzaku/pest.html} which creates a two-dimensional map of the pile-up fraction. In the case of Observation 2b we solve for pile-up by exchanging our source circle for an annulus with an inner region with a radius of $30$ pixels which limits the effect of photon pile-up to $\textless5\%$. Observation 1 was found to be free from any pile-up. The tools \textsc{xisrmfgen} and \textsc{xissimarfgen} were used to create the response and ancillary files respectively. For our analysis we combine the two FI cameras (XIS0 and XIS3) using the FTOOL \textsc{addascaspec} and ignore the BI instrument (XIS1) which has a smaller effective area at 6 keV and is generally less well calibrated. We require a minimum of $20$ counts per bin using \textsc{grppha}. As with the \emph{XMM-Newton} observations, we fit the XIS spectra between 4--10\,keV in \S\ref{Sec:Fe_fits}--\ref{rise_decay} allowing simultaneous fitting with tied parameters and a well calibrated bandpass in a reasonable timescale. For \S\ref{full_test} the bandpass is extended to 0.7--10\,keV and the 1.7--2.4\,keV region is ignored, which like the pn shows strong residuals.

In \S\ref{continuum} and \S\ref{full_test} we also make use of the PIN instrument to extend the bandpass examining the continuum emission. After determining the pointing (XIS or HXD) appropriate response and tuned non-X-ray background (NXB) files were downloaded. We then applied the FTOOL \textsc{hxdpinxbpi} which produces a dead time corrected PIN source spectrum plus the combined PIN background (non X-ray and cosmic). Again a minimum of 20 counts per bin was required and the PIN was fit in the 15--50\,keV bandpass as recommended by the HXD instrument team. The GSO instrument is not used in our analysis.


\subsection{RXTE}
In \S\ref{continuum} and \S\ref{full_test} we use data from the Proportional Counter Array (PCA; \citealt{Jahoda06}) and the High Energy X-ray Timing Experiment (HEXTE; \citealt{Rothschild98}) to extend the continuum bandpass. The data were reduced using HEASOFT software package v6.11 following the standard steps described in the (RXTE) data reduction cookbook\footnote{http://heasarc.gsfc.nasa.gov/docs/xte/data\_analysis.html}. We extracted PCA spectra from the top layer of the Proportional Counter Unit (PCU) 2 which is the best calibrated detector out of the five PCUs, although we added a systematic uncertainty of 0.5\,\% to all spectral channels to account for any calibration uncertainties. We produced the associated response matrix and modelled the background to create background spectra. It has been recently noted (see \citealt{Hiemstra11,Kolehmainen13}) that the cross-calibration between the PCA and EPIC-pn is uncertain, showing an energy-dependant discrepancy at least below $\sim$\,7\,keV. We therefore fit the PCA in the energy range 7--30\,keV for Observations 2 and 3. For Observation 4 the HEXTE was then not operational so we only employed the PCA data, but since the source flux was higher, and the exposure was significantly longer than the other observations, we were able to extend the bandpass up to 50\,keV.

For HEXTE, we produced a response matrix and applied the necessary dead-time correction. The HEXTE background is measured throughout the observation by alternating between the source and background fields every 32s. The data from the background regions were then merged. When possible we used data from both detector A and B to extract source and background spectra. However, from 2005 December, due to problems in the rocking motion of Cluster A, we extracted spectra from Cluster B only. On 2009 December 14, Cluster B stopped rocking as well. From this date, we thus used only PCA data in our analysis. HEXTE channels were grouped by four and fit in the 20--100\,keV band.

For the variability analysis in \S\ref{timing_sec} only the PCA data was used.  We utilised the data modes GoodXenon1\_2s and E\_125us\_64M\_0\_1s depending on the case. Power density spectra (PDS) were computed following the procedure reported in \cite{Belloni06} using stretches 1024s long and PCA channels 0--95 (2--40\,keV).


\section{Analysis and results}
The Fe line region in some of our observations, or contemporaneous ones, has previously been analysed \citep{Tomsick09,Done10,Shidatsu11,Cassatella12,Petrucci14,Kolehmainen13}. Quite simply, one can compare the results from each previous analysis to probe any correlation between the Fe line region and luminosity. However, such a process introduces many issues which could bias results, and hence conclusions. For example, the use of different models; the effects of degeneracies in the models; and the range of values of constant parameters like inclination, are just a few of these potential complications. 

To gain a real insight into evolution throughout an outburst, one must keep the analysis between each observation as consistent as possible. This is the key approach of this work. We fit observations simultaneously and apply the same model to each dataset. Furthermore, we are then able to tie degenerate parameters such as inclination \citep[see Fig.\,3. of][]{Tomsick09}, reducing the effect of such issues. In \S\ref{inclination} we display how degenerate the inclination parameter can be with the disc inner radius. To this end we stress that it is not our aim to present what the exact inner radius value is for each observation. Instead we investigate the relative change of parameters between each observation. This is a crucial step towards an accurate investigation of the Fe line region, and thus the simultaneous analysis of each spectra with tied parameters forms the basis of our investigation (\S\ref{Sec:Fe_fits}--\ref{rise_decay}). For this work we use XSPEC version 12.8.0 and all quoted errors are at the 90\,\% confidence level, unless otherwise stated.

\begin{figure*}
\centering
\epsfig{file=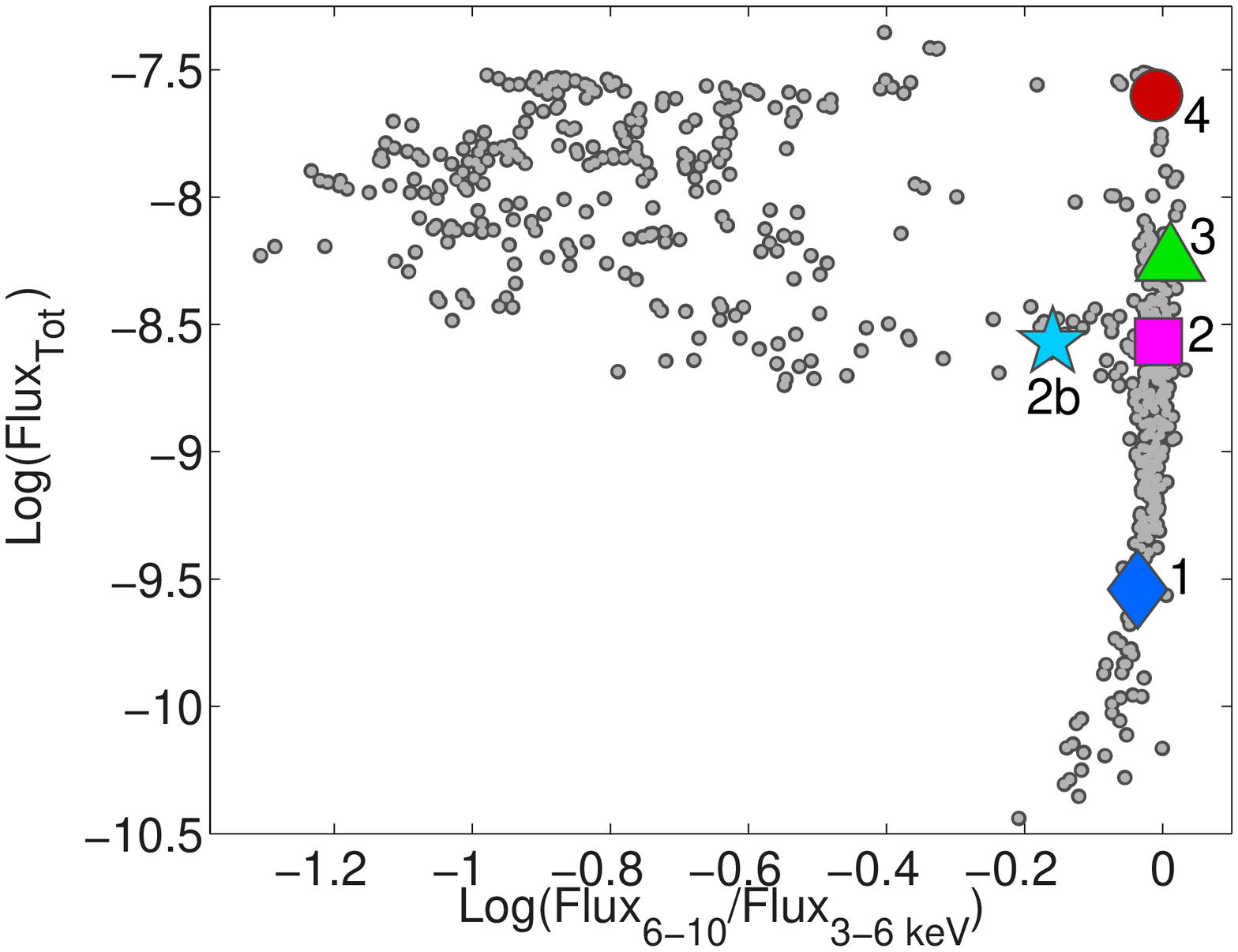, width=0.735\textwidth}
\hspace{-25pt}\epsfig{file=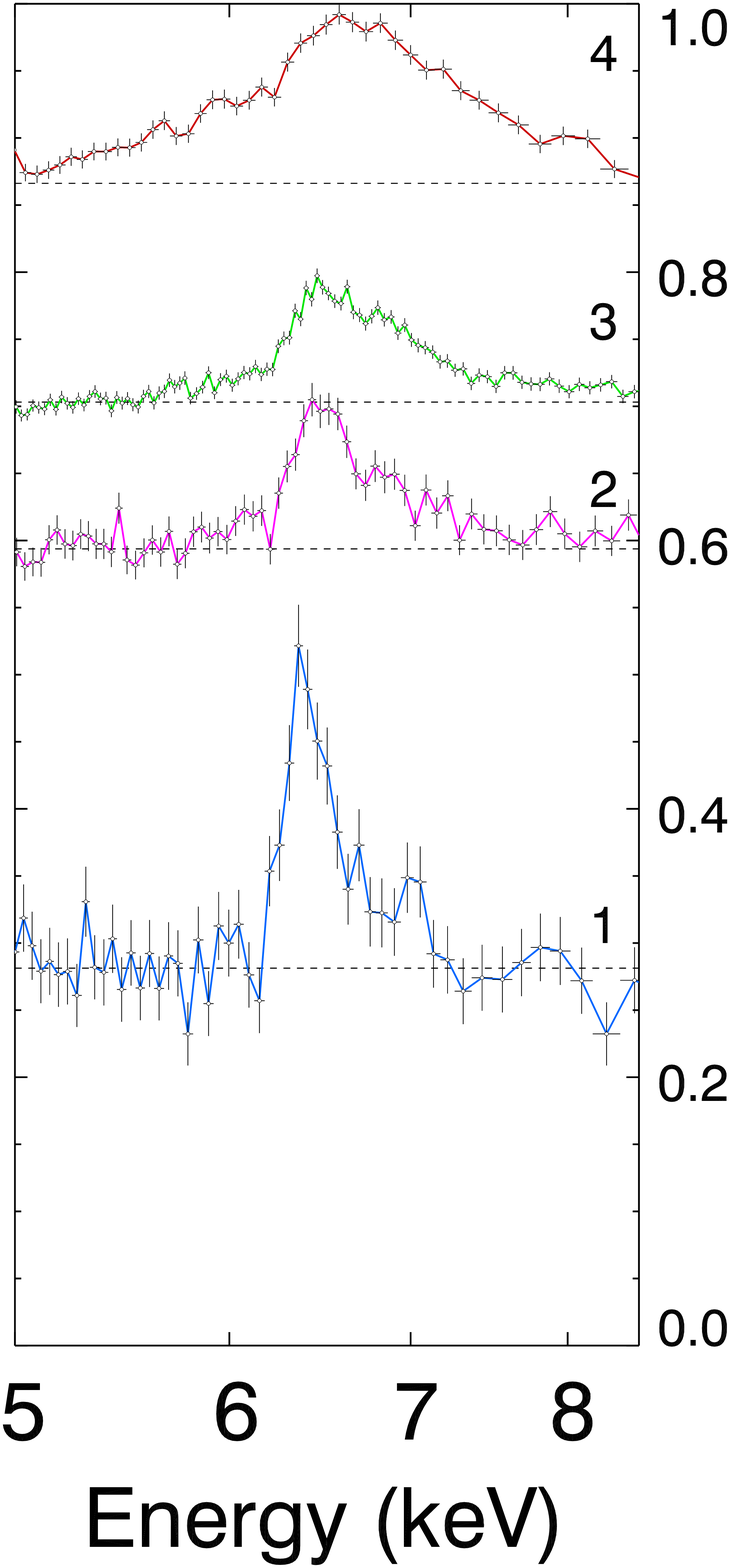, width=0.255\textwidth}
\caption{Left: A Hardness intensity diagram (HID) of all GX 339-4 outbursts monitored by RXTE (1995-2011; grey points). The large coloured symbols indicate the positions of our chosen observations described in Table \ref{observations}: blue/diamond (1), magenta/square (2), green/triangle (3), red/circle (4), and cyan/star (2b). The flux has units of erg cm$^{-2}$ s$^{-1}$. Right: The data/model ratio of continuum fits to the four hard state spectra when the `ignored' energy range (5--8\,keV) in Table \ref{observations} are added (`noticed' in XSPEC) back in, indicating the shape and strength of the Fe line region between observations. Each spectra are shifted arbitrarily to match the left figure and follow the same ratio scale on the y-axis. The data have been re-binned for plotting purposes as well. The cyan HIMS observation (2b) is not shown. This is since it would overlap the Observation 3 ratio and is in a different spectral state to the others. The line profile of Observation 2b is instead displayed in Fig.\,\ref{Fig:HIMS}.}
\label{Fig:HID}
\end{figure*}


\subsection{Our selected models}\label{models}
The X-ray spectrum of BHXRBs can generally be described by three components. Two of these are a thermal blackbody component originating from the accretion disc and a power-law likely due to Compton up-scattered seed photons. The third, known as the `reflection spectrum' \citep{Fabian10}, occurs due to the irradiation of the disc by the up-scattered photons, resulting in, but not exclusively, X-ray fluorescence.  The most prominent signature of this is Fe K$\alpha$ emission due to its high abundance and fluorescent yield.

The hard state itself is dominated by the power-law component, whilst the disc is weak (kT $\textless 0.5$\,keV) above $1$ keV \citep{DiSalvo01,Miller06,Reis10}. Because of this we ignore the region below $4$\,keV above which the disc contribution is negligible, thereby simplifying our spectrum to just 2 components. However, we later check for consistency using a full bandpass, and include a disc emission model, in \S\ref{full_test}. We model the interstellar absorption using the model \textsc{phabs} fixed at $0.5\times 10^{22}$\,cm$^{-2}$. There are two reason for this: firstly, by only fitting above $4$ keV our ability to constrain the column freely is reduced. Secondly, the neutral hydrogen value towards GX 339-4 is well resolved to be within the range $(0.4-0.6)\times10^{22}$\,cm$^{-2}$ \citep{Kong00}. We therefore fix the value of the column the be at the centre of these limits, noting that the effect of such a range above $4$\,keV is negligible. We also note that variability in the absorption has been suggested \citep{Cabanac09}; however, this was found not to be the case when fitting individual photoelectric absorption edges in high-resolution X-ray spectra \citep{Miller09b}.

To model the reflection there are two methods we can use: (1) fit the prominent Fe K$\alpha$ emission line blurred due to relativistic effects  \citep{Fabian89, Laor91} or (2) self-consistently reproduce the entire reflection spectrum for an illuminated accretion disc  \citep{Ross99, Ross05}, whilst blurring relativistically  \citep{Brenneman06, Dauser10}. Each method is extensively applied in the literature and hence we implement them both in our analysis. To model the Fe K$\alpha$ line (method 1) and apply relativistic blurring in method 2 we employ the \textsc{relline} and \textsc{relconv} models of \cite{Dauser10} respectively. For the reflection spectrum we use the code \textsc{xillver}  \citep{Garcia10, Garcia11} which goes beyond the resolution of the most widely used publicly available model \textsc{reflionx} \citep{Ross05}. However, we do include \textsc{reflionx} to allow comparison with previous investigations and ensure our results with \textsc{xillver} are consistent.


\subsection{The Continuum}\label{continuum}

\renewcommand{\arraystretch}{1.2}
\begin{table}
\caption{Results of continuum fits to our five selected observations using the model \textsc{phabs$\ast$powerlaw}, ordered by increasing flux.}
\centering
\begin{tabular}{ccccccc}\hline\hline
Observation			& $\Gamma$				& N$_{\rm PL}$			& Flux 					& $\chi^{2}/\nu$ 			\\ \hline
1					& 1.565$\pm{0.012}$		& 0.011$\pm{0.001}$		& 0.82					& 777/791				\\
2					& 1.437$\pm{0.008}$		& 0.098$\pm{0.001}$		& 8.30					& 714/668				\\
2b					& 1.807$\pm{0.009}$		& 0.144$\pm{0.003}$		& 8.46					& 882/789				\\
3					& 1.476$\pm{0.003}$		& 0.186$\pm{0.001}$		& 15.05					& 812/666				\\
4					& 1.647$\pm{0.004}$		& 0.919$\pm{0.006}$		& 59.61					& 765/658				\\\hline
\end{tabular}
\tablefoot{An additional \textsc{smedge} is added to Observation 4 to account for the significant residual beyond $8$\,keV. The unabsorbed flux is in units of $\times 10^{-10}$\,erg cm$^{-2}$ s$^{-1}$ and calculated in the 0.5--10\,keV energy range using the model \textsc{cflux}.}
\label{table_continuum}
\end{table}

To begin, we fit each observation individually with the Fe line region (5--8\,keV) removed in order to accurately estimate the continuum emission upon which the reflection is superimposed. At this point we want to test two properties of the continuum: (1) whether an absorbed power-law is an acceptable description of the continuum, and (2) that our description of the continuum is consistent with the later analysis when the reflection is modelled. It is imperative that our continuum is correctly modelled since the shape of the reflection spectrum is strongly dependent upon it, and hence we use this initial analysis as a sanity-check for the later sections. To this end, the motivation for fitting each observation individually in this section is that it allows us to extend our broadband coverage. Our main focus, the EPIC-pn and XIS cameras, will only be covering the 4--10\,keV region to constrain the continuum for the majority of this investigation. We add simultaneous PCA-HEXTE observations to the EPIC-pn and make use of the PIN on-board Suzaku to extend the coverage up to 100\,keV and 50\,keV respectively. A constant normalisation factor must be allowed between each detector; however, the only one well calibrated enough to be fixed is between the XIS and PIN (both of our Observations were taken in the XIS nominal position, and hence we fix the constant to be 1.16). The cross-normalisation between the PCA, HEXTE and pn are all uncertain, so the PCA constant was fixed to be 1 and the pn and HEXTE datasets are allowed a free constant. The fitted constants were $\sim$0.85 and $\sim$0.8 for the pn and HEXTE, respectively.

We fit a single absorbed power-law to the broadband spectrum above 4\,keV, tying the photon index and normalisation parameters between detectors whilst allowing a constant to float between them as described before. Our continuum model is found to be a good fit to all of the observations (Table \ref{table_continuum}) and no cut-off is required at high energies, although this may exist beyond our upper energy limit.  A significant smeared edge is required for Observation 4 suggesting a large amount of reflection is present. The majority of residuals lie beyond 10 keV, the likely source of which is a Compton `hump' from the reflection of hard X-rays by the cool accretion disc. Interestingly, there is no obvious edge or Compton hump present in data--model ratios of Observations 1 and 2, suggesting that the level of reflection has increased in the higher luminosity observations. However, since 1 and 2 are the lowest flux observations, these features may be masked by the lower signal-to-noise ratio. Given the good description of the continuum by this model, we use it as the base continuum in the later sections unless stated otherwise. We do not include the smeared edge applied to Observation 4 in the later sections. This is because it may interfere with the Fe line fitting, and using the same continuum model allows the fairest comparison of each observation. In addition, the edge is self-consistently calculated in the reflection models \textsc{reflionx} and \textsc{xillver} (\S\ref{full_ref} onwards), and is hence not required. 

For now we only consider the hard state spectra. We display the continuum data to model ratio with the Fe line region added back into the plot (\ie `noticed' in XSPEC) for the hard state observations in Fig.\,\ref{Fig:HID}. Immediately one can see a distinct evolution of the Fe line region. At higher luminosities the profile extends further in both the red and blue wings. Additionally, the peak appears to shift to higher energies. A higher spin and inclination will broaden the red wing and blueshift the peak respectively. However, these two parameters will be constant between observations, so we can rule out their influence.

Two remaining variables can increase the profile's extension to lower energies. Were the inner disc radius to change, specifically to extend closer to the BH, the relativistic effects would increase and we would observe an increase in the broadening of the red wing. Additionally, the emissivity can generate this effect. More centrally concentrated emission would mean a larger contribution to the profile by region experiencing stronger relativistic effects, so increased red wing broadening. However, we do not expect the emissivity to vary so intensively between observations given that previous investigations have yielded $q\sim3$ (the emissivity is defined to scale as $r^{-q}$; \citealt{Miller06b,Reis08,Reis10,Done10,Shidatsu11}). Furthermore, if the illuminating source (the corona) had changed between these observations, this would only strongly affect the value of $q$ corresponding to the very inner regions of the disc (\textless $2\,r_{\rm g}$). As noted by \cite{Fabian12}, even for the high spin estimated for GX 339-4 \citep{Reis08}, $q$ should not deviate significantly away from $3$ at the ISCO. Thus inner radius variation should dominate the different red wings that we observe. The shift in the peak of the profile is almost certainly due to an increase in the disc ionisation. A more ionised disc means emission from higher rest energies, and hence a shift in the peak of the line profile. Also, larger ionisation results in increased Compton scattering and emission from multiple ionisation stages which can both contribute an overall broadening of the profile  (\citealt{Garcia09};  see also \citealt{Garcia10} and \citealt{Garcia11}), thus extending the blue wing.

Therefore, visual inspection of Fig.\,\ref{Fig:HID}, just by eye, implies that at higher luminosities in the hard state the accretion disc is extending closer to the BH and becoming more ionised.

\renewcommand{\arraystretch}{1.5}
\begin{table*}
\centering
\caption{Results from fitting the four hard state observations simultaneously with the \textsc{relline} line profile (top section) and a blurred reflection model \textsc{relconv$\ast$xillver} (middle section), \textsc{relconv$\ast$reflionx} (bottom section).}
\begin{tabular}{cc cccc}\hline\hline
Model				& Parameter					& 1					& 2					& 3					& 4					\\\hline
	
\textsc{powerlaw}		& $\Gamma$					& 1.59$\pm{0.02}$		& 1.42$\pm{0.01}$		& 1.47$\pm{0.01}$		& 1.64$\pm{0.01}$		\\
					& N$_{\rm PL}$					& 0.011$\pm{0.001}$	& 0.096$\pm{0.001}$	& 0.18$\pm{0.01}$		& 0.91$\pm{0.01}$		\\
\textsc{relline}			& E (keV)						& 6.46$\pm{0.02}$		& 6.55$^{+0.02}_{-0.03}$	& 6.94$^{+0.02}_{-0.03}$	& 6.97$_{-0.01}$	\\
					& $\theta$ ($^{\circ}$)			& \multicolumn{4}{c}{18$\pm0.1$}														\\
					& $r_{\rm in}$ ($r_{\rm isco}$)		& 79$^{+68}_{-26}$		& 36$^{+11}_{-8}$		& 3.7$^{+0.2}_{-0.1}$	& 3.0$\pm{0.2}$	\\	
					& N$_{\rm L}$ ($10^{-4}$)			& 0.37$\pm{0.05}$		& 2.61$\pm{0.22}$ 		& 7.85$^{+0.30}_{-0.25}$	& 59.3$\pm{3.0}$	\\
					& E.W. (eV)					& 61$^{+10}_{-7}$		& 40$^{+4}_{-5}$		& 75$^{+3}_{-5}$		& 154$^{+13}_{-6}$		\\
\multicolumn{2}{c}{$\chi{^2}/\nu$}						& \multicolumn{4}{c}{6608/5100}														\\
\hline
\textsc{powerlaw}		& $\Gamma$					& 1.72$^{+0.15}_{-0.03}$	& 1.44$\pm{0.01}$		& 1.47$\pm{0.01}$		& 1.60$\pm{0.01}$		\\
					& N$_{\rm PL}$					& 0.013$\pm{0.002}$	& 0.087$\pm{0.004}$	& 0.16$\pm{0.01}$		& 0.37$^{+0.06}_{-0.04}$	\\
\textsc{relconv}			& $\theta$ ($^{\circ}$)			& \multicolumn{4}{c}{42$^{+11}_{-6}$}													\\
					& $r_{\rm in}$ ($r_{\rm isco}$)		& \textgreater 344		& 295$^{+131}_{-163}$		& 137$^{+71}_{-32}$	& 67$^{+60}_{-23}$		\\
\textsc{xillver}			& $\log(\xi)$					& 1.52$^{+0.32}_{-0.40}$	& 2.58$^{+0.03}_{-0.04}$	& 2.61$^{+0.02}_{-0.01}$	& 2.88$^{+0.03}_{-0.02}$	\\
					& N$_{\rm R}$ ($10^{-6}$)		& 6.50$^{+25.6}_{-3.91}$	& 3.31$^{+0.27}_{-0.24}$	& 4.82$^{+0.15}_{-0.14}$	& 20.6$^{+0.09}_{-0.07}$	\\
					& $RF$						& 0.13$\pm{0.05}$		& 0.13$\pm{0.03}$		& 0.14$\pm{0.01}$		& 1.29$\pm{0.25}$	\\
\multicolumn{2}{c}{$\chi{^2}/\nu$}						&\multicolumn{4}{c}{5467/5100}														\\
\hline
\textsc{powerlaw}		& $\Gamma$					& 1.66$\pm{0.03}$		& 1.45$\pm{0.01}$		& 1.49$\pm{0.01}$ 		& 1.53$^{+0.03}_{-0.07}$\\
					& N$_{\rm PL}$					& 0.012$\pm{0.001}$	& 0.097$\pm{0.001}$	& 0.18$\pm{0.01}$		& 0.44$^{+0.12}_{-0.13}$	\\
\textsc{relconv}			& $\theta$ ($^{\circ}$)			& \multicolumn{4}{c}{36$^{+3}_{-6}$}													\\
					& $r_{\rm in}$ ($r_{\rm isco}$)		& \textgreater 321		& 419$^{+7}_{-277}$		& 54$^{+9}_{-22}$		& 31$^{+18}_{-9}$		\\
\textsc{reflionx}			& $\log(\xi)$					& 2.06$^{+0.08}_{-0.34}$	& 2.38$^{+0.02}_{-0.01}$ 	& 2.47$\pm{0.03}$		& 3.36$\pm{0.01}$		\\
					& N$_{\rm R}$ ($10^{-6}$)		& 2.47$\pm{0.31}$		& 8.38$^{+1.00}_{-0.51}$	& 8.44$^{+0.84}_{-0.93}$	& 6.57$^{+0.38}_{-0.35}$	\\
					& $RF$						& 0.10$\pm{0.03}$		& 0.06$\pm{0.03}$		& 0.06$\pm{0.01}$		& 0.73$\pm{0.21}$	\\
\multicolumn{2}{c}{$\chi{^2}/\nu$}						&\multicolumn{4}{c}{5482/5100}														\\\hline
\end{tabular}
\tablefoot{The photon index in the reflection models is tied to that of the continuum power-law and the Fe abundance is assumed to be solar. The emissivity and spin parameters are fixed to be $r^{-3}$ and 0.9 respectively, whilst the outer radius of the disc is fixed to be 1000\,$r_{\rm g}$. We calculate the reflection fraction ($RF$) as the ratio of the flux from the reflected emission and the power-law continuum flux. Both are calculated using {\sc cflux} in the 4--10\,keV band. If an upper or lower limit is not shown this indicates that the parameter has reached a hard limit. In some cases the inner radius parameter has reached the largest tabulated value in the model ($1000\,r_{\rm g}$), therefore we only present the lower limit in this case.}
\label{fits}
\end{table*}


\subsection{The Fe line region}\label{Sec:Fe_fits}
\subsubsection{Line modelling}\label{line_model}
\begin{figure}
\centering
\centerline{\epsfig{file=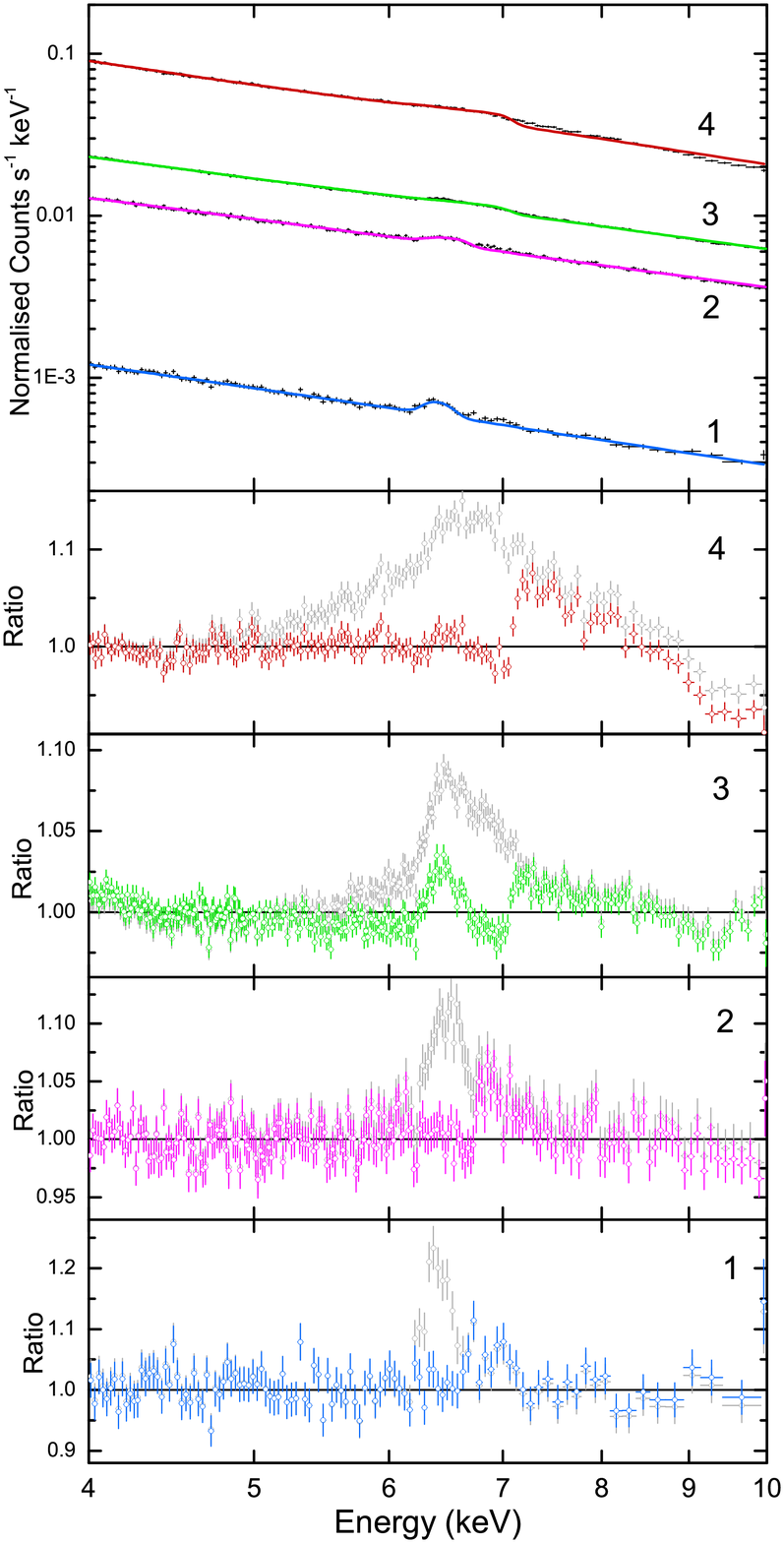, width=0.5\textwidth}}
\caption{Our four hard state datasets simultaneously modelled using \textsc{relline} for the Fe K$\alpha$ emission line. The top panel shows the resultant best fit with the remaining panels displaying the model residuals for each observation using the same colour scheme as in Fig.\,\ref{Fig:HID}. Additionally the model residuals from Fig.\,\ref{Fig:HID} are overlaid in grey. These correspond to the absorbed power-law continuum ({\sc phabs$\ast$powerlaw}) before the Fe line is fitted, as described in $\S$\ref{continuum}. This therefore indicates how the \textsc{relline} model is fitting the overall shape of the Fe band. All of the spectra have been re-binned for the purposes of plotting. Note as well that the line profile of Observation 4 is slightly different to Figure\,\ref{Fig:HID}, which included the \textsc{smedge} model.}
\label{Fig:relline}
\end{figure}

To gain a more accurate description of how the disc is evolving we next model the Fe line region, and we begin by fitting the Fe K$\alpha$ line. We employ the model \textsc{relline} \citep{Dauser10}, which assumes an intrinsic zero width emission line transformed by the relevant relativistic effects. The rest energy of the emission is fitted freely within the limit 6.4--6.97\,keV, which represents the range of emission from neutral to H-like Fe. For our initial analysis we focus upon the four hard state observations in Table \ref{observations} to probe how the reflection evolves just within this state. We fit these simultaneously, tying the inclination between each observation and fix the emissivity and spin to be $r^{-3}$ and $0.9$ respectively, and the outer radius of the disc is fixed at $1000\,r_{\rm g}$. We adopt 0.9 for the spin parameter since this is the upper limit found by \cite{Kolehmainen10} using the continuum method. It is also close to values obtained previously through reflection fitting \citep{Reis08, Miller08}. We discuss this impact of this choice in \S\ref{spin_discuss}. In order to fit the four epochs simultaneously we now only use the EPIC-pn and combined XIS datasets, and fit these between 4--10\,keV. We use the absorbed power-law continuum described in \S\ref{continuum}, applying those results as the initial parameter values. Our results are presented in Table \ref{fits}. The joint best-fit is reasonably good ($\chi^2/\nu$ = 1.30), but significant residuals are evident, particularly in the Fe K band (Fig.\,\ref{Fig:relline}). The feature around 7 keV in Observation 1 is likely to be Fe K$\beta$ emission, consistent with the near-neutral Fe K$\alpha$ fit of 6.46$\pm{0.02}$\,keV. A similar residual is also seen in Observation 2 but below 7 keV, which possibly arises from emission from a higher ionisation stage (H-like at 6.97\,keV) or Compton scattering broadening the profile. However, Fe K$\beta$ emission cannot be ruled out. The effect of multiple ionisation stages is also seen in Observation 3 where a feature at 6.4\,keV remains, most likely due to neutral emission from the outer disc, and a broad residual is seen beyond 7\,keV as scattering is not modelled. In Observation 4 the disc is likely to be more ionised removing the former issue, but the latter becomes even more significant as a result. A large edge is also clearly present\footnote{We do not include the \textsc{smedge} model like in \S\ref{continuum} in order to fairly compare the four observations.}.

Our results indicate consistently that the inner accretion disc is more recessed at lower luminosities. We see evolution from 79\,$r_{\rm isco}$ to 3\,$r_{\rm isco}$ over two magnitudes of L$_{\rm Edd}$ (Table \ref{fits}). However, we stress that our recorded values may not be a true indication of the inner disc radius as it is possible degeneracies in the model could skew its accuracy. Furthermore, we can see from Fig.\,\ref{Fig:relline} that the reflection is not well fitted by a single line. The key result here is the strong relative trend, which we can be confident about, of a disc extending closer to the BH at higher luminosities. This confirms the first of our predictions using Fig.\,\ref{Fig:HID}. Additionally, we note the rest energy of emission also increases with luminosity, approaching the upper limit of the allowed range ($6.97$\,keV; H-like Fe K$\alpha$ emission) for Observations 3 and 4. This, therefore, predicts that the disc surface layers are more ionised at higher luminosities, adding weight to the second and third of our predictions by eye. The tied inclination parameter is found to be 18$^{\circ}$, which is consistent with previous analysis using a single Fe line \citep{Miller06b,Reis10,Done10}. We note, however, that an inclination lower than $\sim40^{\circ}$ looks unfeasible since it will result in BH mass of $\textgreater$\,20\,$M_{\sun}$ (assuming the constraints reported in \citealt{Hynes03} and \citealt{Munoz08}), although the inner disc may not be aligned with the binary inclination \citep{Maccarone02}.

The equivalent width (EW) of the lines in the rise data increases with luminosity, indicating a larger area of reflection ({\it i.e.} a smaller inner radius). However, Observation 1, which is in decay, does not fit this trend. It is likely, though, that this is the only profile well described by one ionisation stage, and hence would yield a larger value compared to other observations, which instead leave significant residuals in their fit (Fig. \ref{Fig:relline}). We therefore believe the reflection fraction ($RF$) calculated in \S\ref{full_ref} to be a more accurate method to draw any conclusions from the relative flux between components.


\subsubsection{Self-consistent reflection fitting}\label{full_ref}
As noted in the previous section, and displayed in Fig.\,\ref{Fig:relline}, although the fit was statistically satisfactory when line modelling was applied, there were residuals remaining in the Fe band. This suggests that the reflection is more complex than just Fe K$\alpha$ emission at one ionisation stage. Such a shortcoming is likely to arise due to Comptonisation and emission from regions of the disc with different physical conditions. Both would generate further broadening and structure in the resolved emission line, and these mechanisms are not taken into account by a Fe line model. Therefore, the model will either be artificially broadened in order to fit the profile, for example by over-predicting the relativistic effects, or leave the additional broadening not modelled, and consequently residuals in the spectra.

\begin{figure}
\centering
\centerline{\epsfig{file=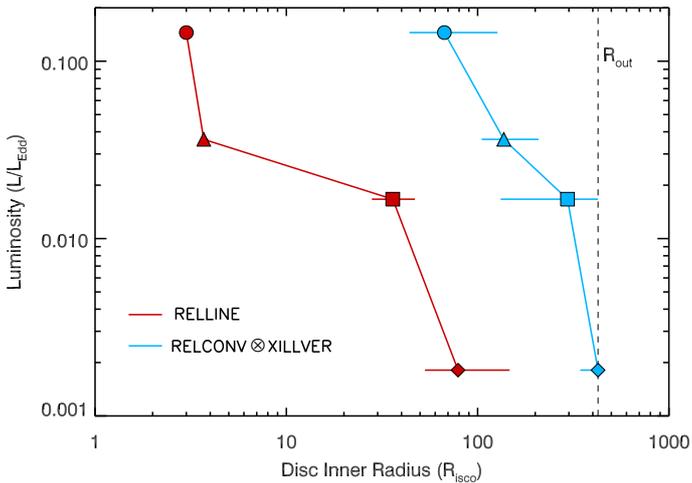, width=0.5\textwidth}}
\caption{Evolution of the estimated inner radii (x-axis; Table \ref{fits}) as a function of luminosity (y-axis; assuming a BH mass and distance of $8$\,$M_{\sun}$ and $8$\,kpc respectively), clearly showing that the inner radius decreases as the source luminosity rises. The red and blue lines refer to fits with \textsc{relline} and {\sc relconv$\ast$xillver} respectively. The black dashed line indicates the disc outer radius, which was fixed to be $1000\,r_{\rm g}$.}
\label{Fig:rin_lum}
\end{figure}

We therefore compare our results from line modelling to that of relativistically blurred reflection from an ionised disc. The model we employ, on top of our standard continuum, is \textsc{relconv$\ast$xillver} \citep{Dauser10,Garcia10,Garcia11}. Again, we use a single absorbed power-law as the underlying continuum, using the results in \S\ref{continuum} as the initial parameter values, but otherwise fitted freely. The inclination is tied between each observation, and the emissivity and spin are fixed to be $r^{-3}$ and $0.9$ respectively, whilst the outer radius of the disc is fixed at $1000\,r_{\rm g}$. The input photon index in the reflection model is tied to that of the power-law, and solar abundances are assumed. The fit is improved ($\chi^2/\nu$ = 1.07; Table \ref{fits}), and we record that all four observations are expected to harbour a significantly more truncated disc relative to the values obtained from modelling the Fe K$\alpha$ line only. The evolution of the inner radius is found to be less significant statistically than for method 1; however, we still find strong evidence for evolution between Observations 4 with 1 and 2, and Observation 3 with 1. Figure\,\ref{Fig:rin_lum} displays the evolution of the inner radius against source luminosity, and emphasises for both methods how the disc moves inwards as the luminosity rises.

The much larger values recorded for the inner radii can be driven by two effects. A much larger inclination is found (42$^{\circ}$ vs. 18$^{\circ}$) and the smaller doppler shifts at low inclinations would require a smaller inner radius to recover the same profile width (\citealt{Done10}; see also \S\ref{inclination}). However, as discussed in the previous section, the much lower value found by the single Fe line method is highly unlikely given the constraints on the binary parameters of GX 339-4. Thus, the inner radii values found for the self-consistent reflection method are more likely to be correct, and we discuss the accuracy of the resolved inner disc radius further in \S\ref{rtest}. We investigate the effect of the inclination parameter further in \S\ref{inclination}.

\begin{figure*}
\centering
{\epsfig{file=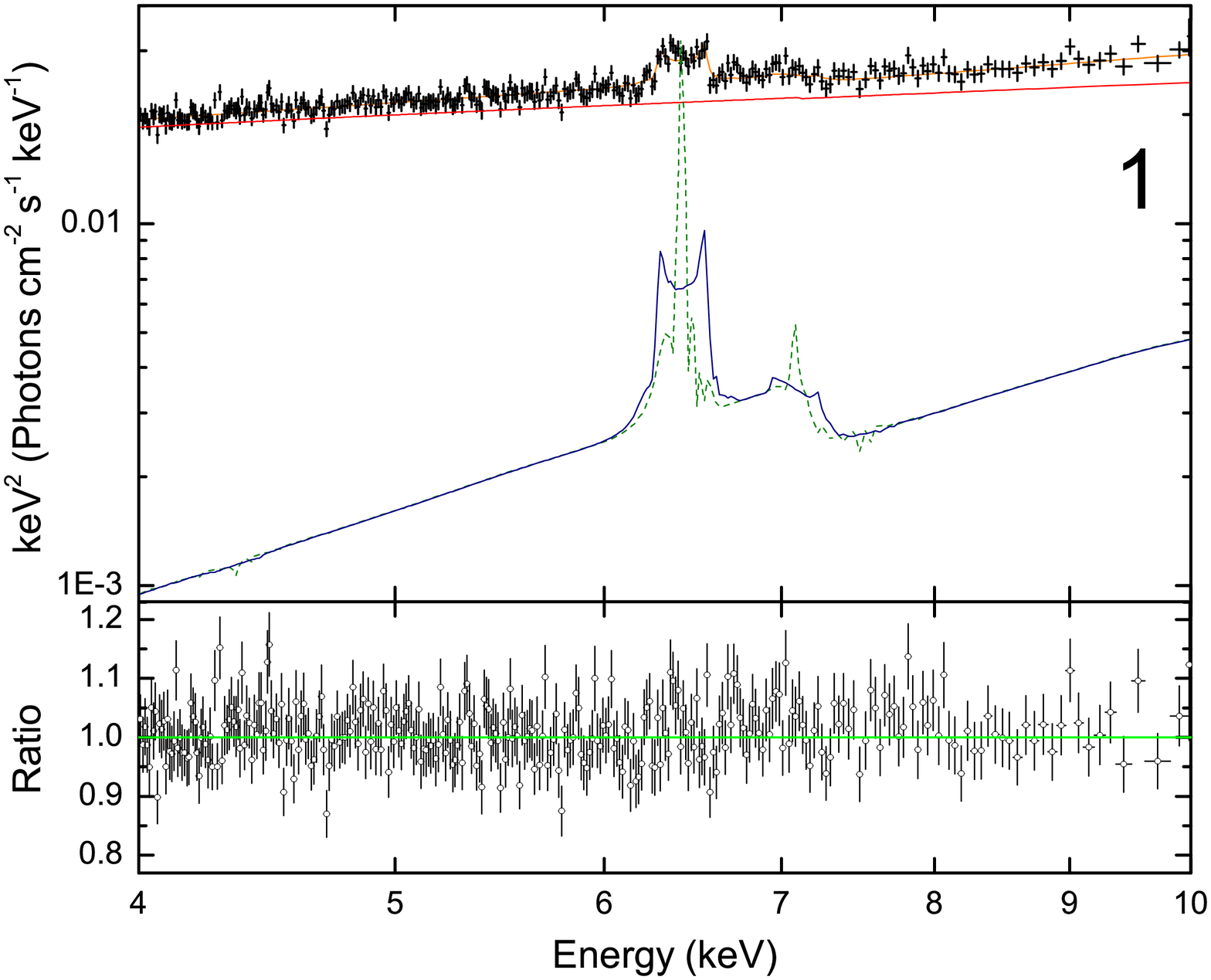, width=0.45\textwidth}}
\hspace{25pt}{\epsfig{file=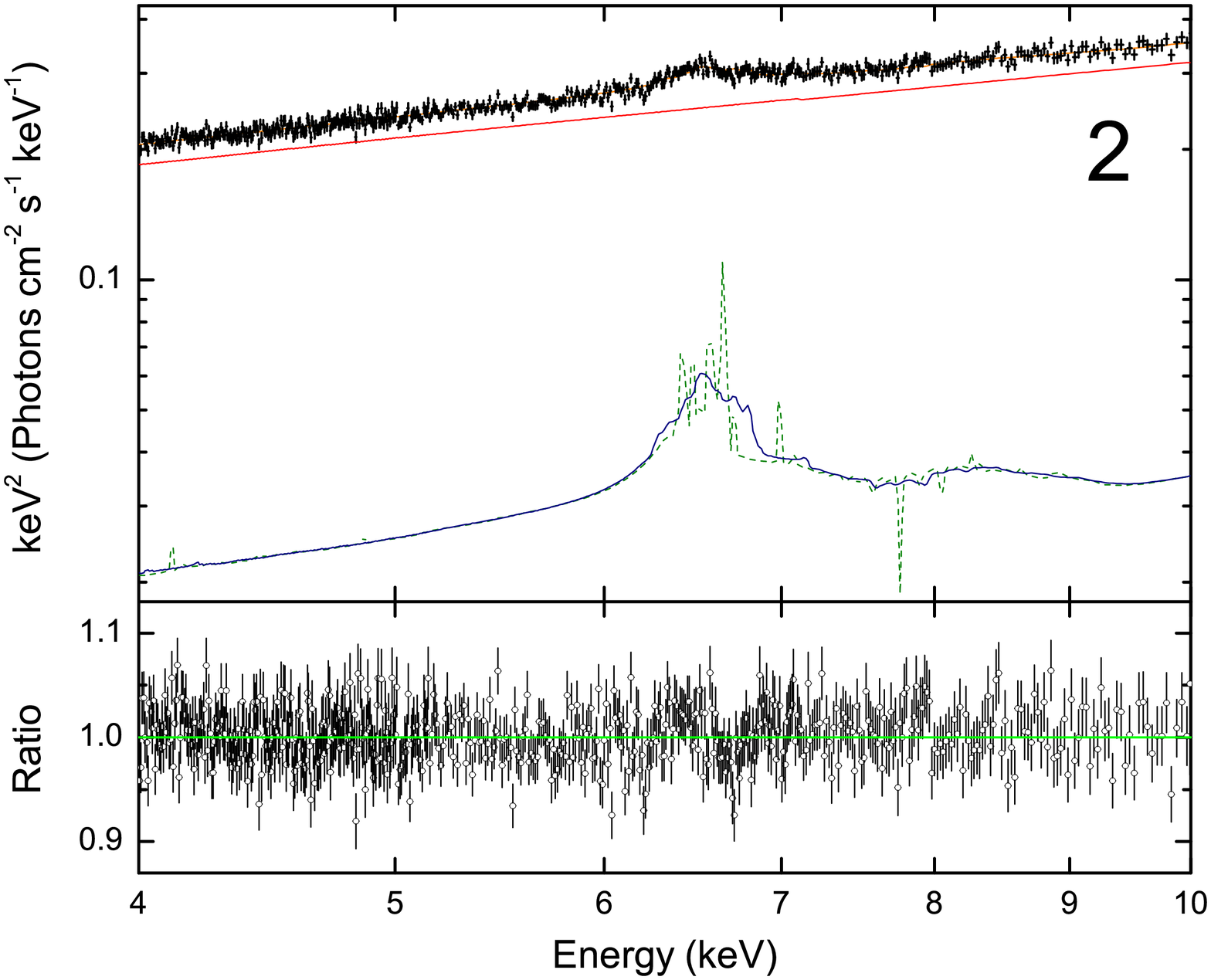, width=0.45\textwidth}}
{\epsfig{file=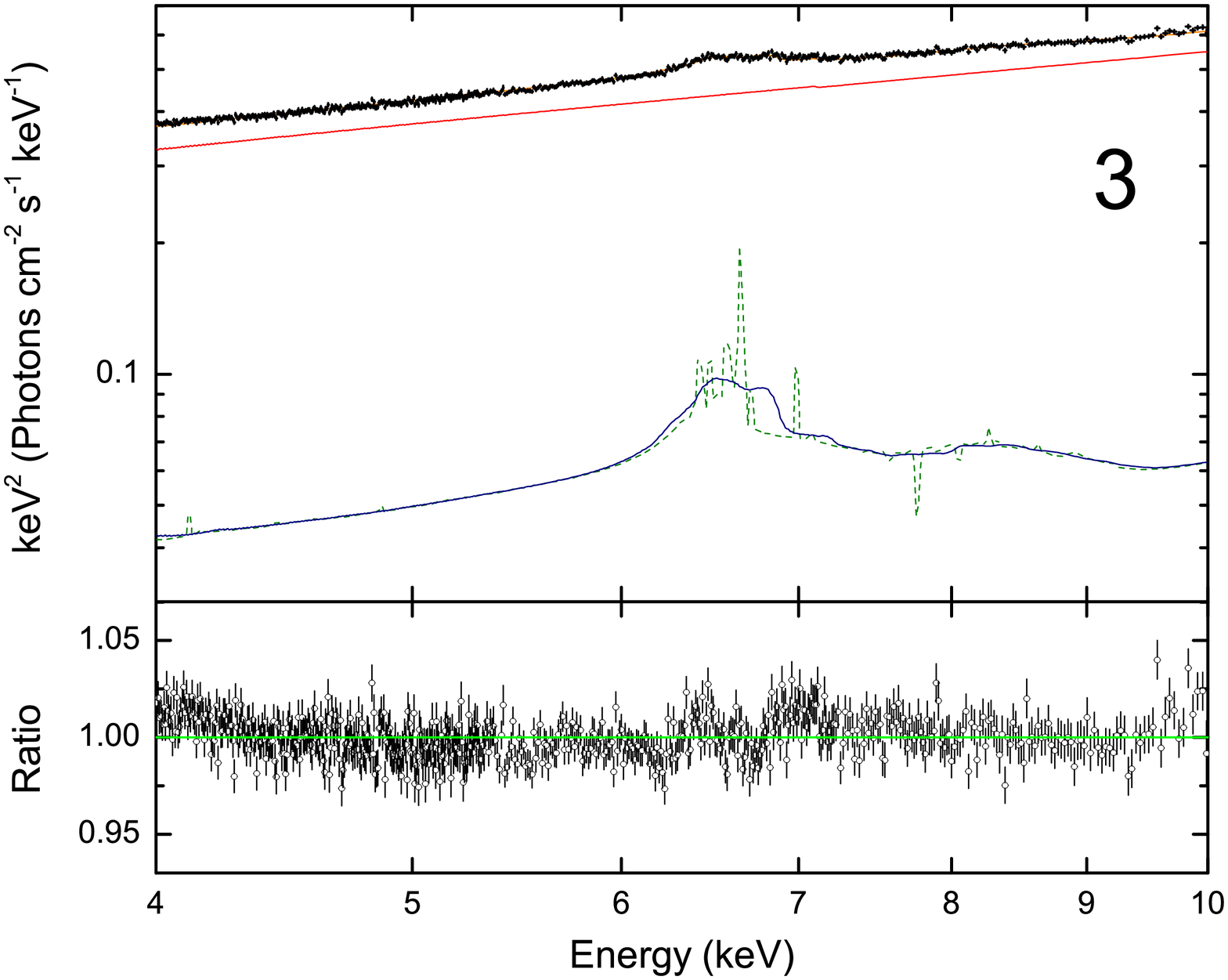, width=0.45\textwidth}}
\hspace{25pt}{\epsfig{file=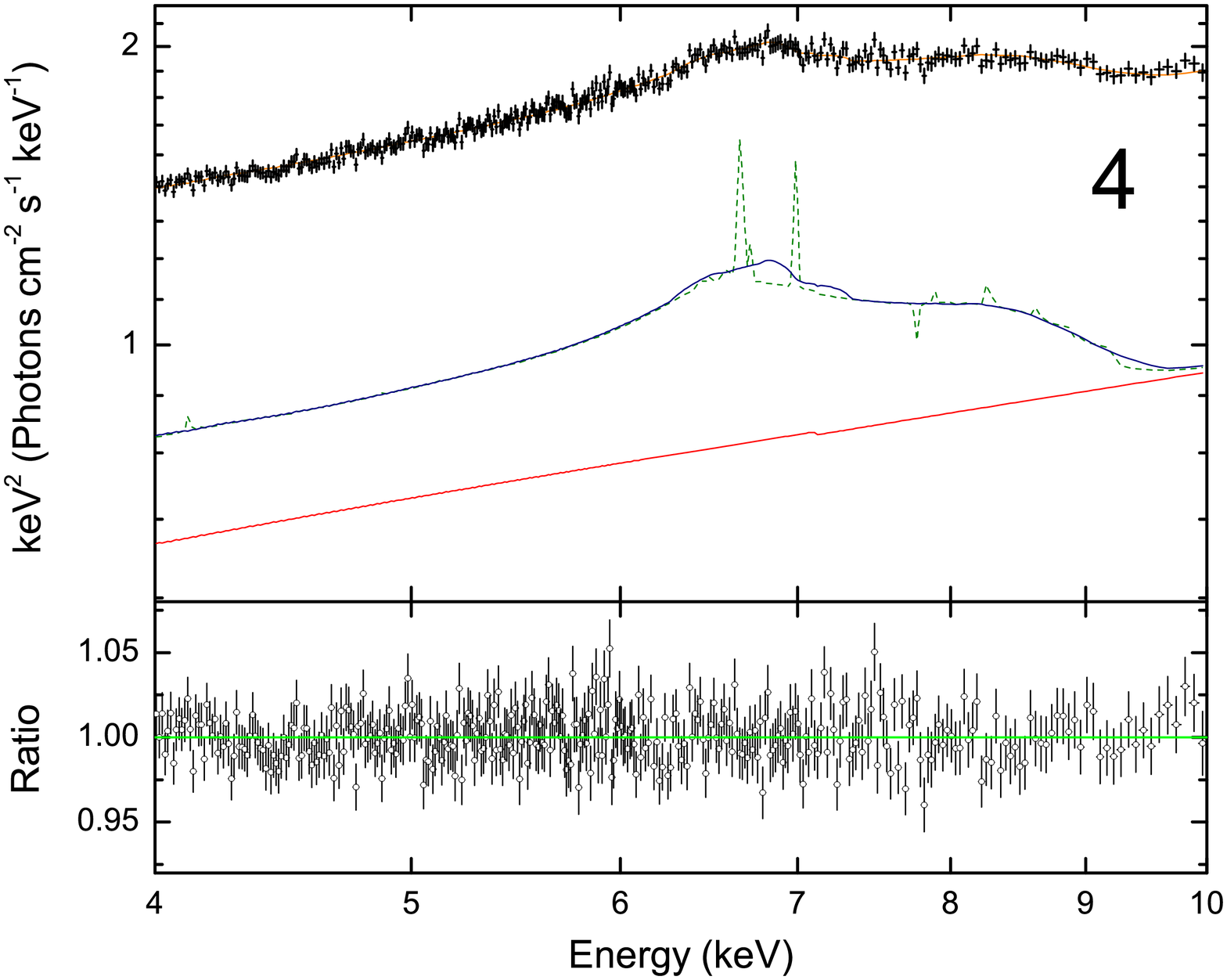, width=0.45\textwidth}}
\caption{The unfolded spectral fit and data/model ratios of the simultaneous fitting of the four hard state observations in Table \ref{observations}, using the blurred reflection model \textsc{relconv$\ast$xillver}. The up-scattered disc emission is indicated by the red lines and the composite model is overlaid in orange. We present the reflection in two ways: the blue solid lines show the blurred reflection as fitted, whilst the green dashed lines display the reflection spectrum when the relativistic blurring (\textsc{relconv}) is removed. Please note that the green dashed lines are not part of the fit, they are purely to allow the evolution of the profile due to ionisation and relativistic effects to be viewed separately. All of the spectra have been re-binned for the purposes of plotting.}
\label{Fig:EEUF}
\end{figure*}

Additionally, the ionisation parameter, $\xi=4\pi F/n$ (where $F$ is the total illuminating flux, and $n$ the electron number density), will lead to a broader profile through increased Comptonisation and multiple ionisation stages. We find this spans over a magnitude in range, yielding reasonable values in comparison to previous work. Also present is a trend to higher ionisation with luminosity, consistent with what one would expect: as the accretion rate increases the disc will become warmer, and hence more ionised. Additionally, the more luminous hard component, driving the vertical track on the HID, will irradiate the disc more and more, again increasing the ionisation of the surface layers responsible for the reflection spectrum. In Fig.\,\ref{Fig:EEUF} we can see how the ionisation stage evolves through the green lines, which are the reflection spectrum when the relativistic blurring has been removed. Mainly neutral emission and the significance of Fe K$\beta$ is seen for Observation 1, whilst Observation 4 requires only H and He-like Fe K$\alpha$ emission from an almost fully ionised disc, with a hint of Fe K$\beta$ arising from recombination. Intermediate to this in luminosity, Observations 2 and 3 indicate a moderately ionised disc with emission from a range of stages.

The flux of each component is calculated using the convolution model \textsc{cflux}, keeping all parameters free and tied as before. The reflection fraction ($RF$), which we define at the ratio of the flux from the reflection emission and power-law continuum flux\footnote{Traditionally, in models such as \textsc{pexrav} and \textsc{pexriv} \citep{Magdziarz95}, $RF=\Omega/2\pi$, where $\Omega$ is the solid angle subtended by the disc beneath the illuminating source.}, also shows some increase with luminosity. Although strong conclusions are limited by our chosen bandpass, if the accretion disc is moving inwards it presents a larger solid angle as seen by the illuminating source, therefore increasing the flux due to reflection. This therefore agrees with our evolution of the inner disc, and displays consistency with other investigations of the hard state (see \eg \citealt{Gilfanov99,Plant14}). The values of the power-law photon index show good consistency with those found over a larger bandpass (Table \ref{table_continuum}; \S\ref{continuum}), although some variation is expected as the reflection spectrum is then accounted for with \textsc{xillver} and \textsc{reflionx}. Nevertheless, this indicates that the reflection spectrum is being fitted correctly.

This result stresses the importance in taking into account non-relativistic broadening mechanisms when fitting a line profile. In fact, for Observations 1 and 2 we only find a lower limit for the inner radius, making them consistent with requiring no relativistic broadening at all at the 90\% confidence level. Upon reaching the limits of the model in terms of radii, the line profile shape becomes dominated by ionisation rather than relativity, making the inner radius parameter difficult to constrain. The consistency of the evolution we find between observations brings strong evidence that the inner accretion disc can be substantially truncated in the hard state, even at relatively high luminosities (up to $\sim$0.15\,L$_{\rm Edd}$\footnote{Assuming a BH mass of 8\,$M_{\sun}$ at a distance of 8\,kpc}).

We also repeated our analysis replacing {\sc xillver} with the reflection model {\sc reflionx} \citep{Ross05}, finding similar evolution. For this analysis we applied the same fixed parameters as used for \textsc{xillver}. We see that the inner radius parameter follows the same trend, although having smaller values for each observation. The inclination is found to be less ($\sim$36$^{\circ}$), and this decrease is likely to drive the trend to lower values of $r_{\rm in}$ in the same way we proposed for the line only method. Observations 1 and 2 are still consistent with requiring no relativistic broadening at the 90\,\% confidence level. The ionisation parameter finds different values, but this is likely due to differences between the two models, and the trend of higher ionisation with increasing luminosity is still very clear.

\subsection{Markov Chain Monte Carlo analysis}\label{Sec:MCMC}
The three different models we have applied have unanimously observed the same trends; however, applying $\chi^2$ fitting to such a large number of free parameters (21; Table \ref{fits}) across four datasets brings with it the considerable possibility that a local minimum is mistaken for the global best fit. Furthermore, as discussed previously, the precise resolution required, and the subtle effects that parameters have, make reflection fitting highly susceptible to degeneracies, which can force the parameter space a fit occupies. In order to assess this possibility we apply a Markov Chain Monte Carlo (MCMC) statistical analysis to our results with \textsc{xillver}. This technique has recently been successfully used by \cite{Reynolds12} to constrain the black hole spin of NGC 3783 which also required a large number of free parameters.

Starting a random perturbation away from our best fit (Table \ref{fits}) we run four 55,000 element chains, of which the first 5000 elements of each chain are discarded (`burnt'). The chain proposal is taken from the diagonal of the covariance matrix calculated from the initial best fit. For this we assume probability distributions which are Gaussian, and a rescaling factor of 10$^{-3}$ is applied. We can use the resultant 200,000 element chain to determine the probability distribution for each parameter and compare to the results found through the standard XSPEC \textsc{error} command. For Observations 1 to 4 the best fit inner radii and MCMC-determined 90\,\% confidence intervals are $\textgreater302$, $295^{+125}_{-120}$, $137^{+84}_{-24}$ and $67^{+52}_{-15}$ respectively, and hence show excellent consistency with those quoted in Table \ref{fits}.

\begin{figure*}
\centering
{\epsfig{file=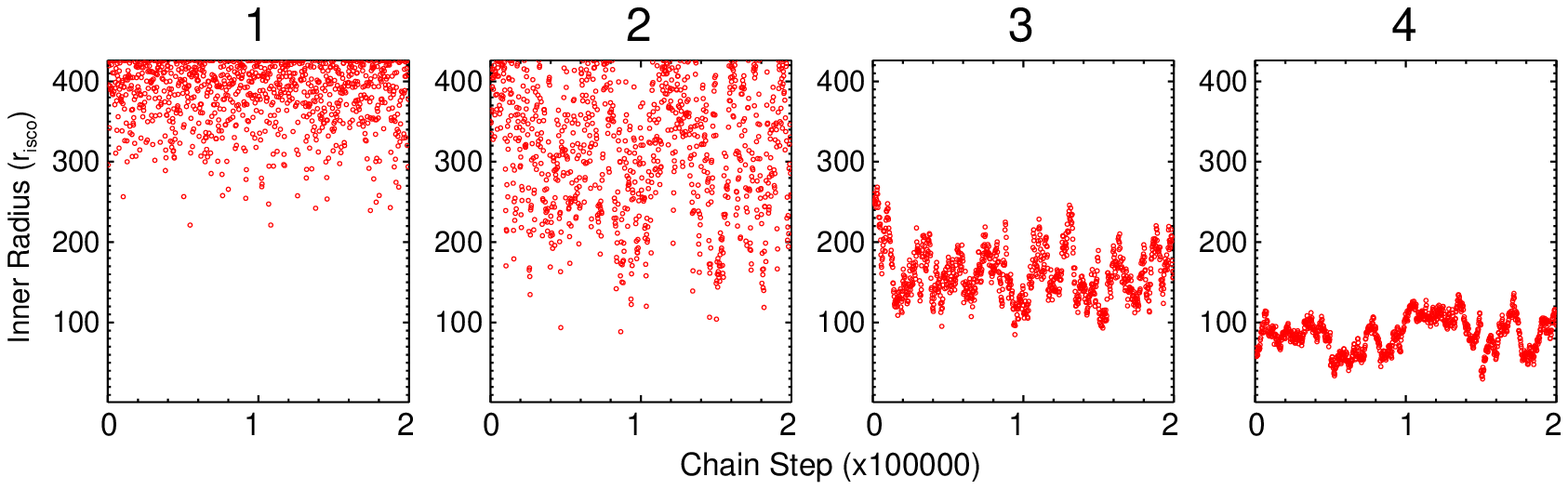, width=0.95\textwidth}}
{\epsfig{file=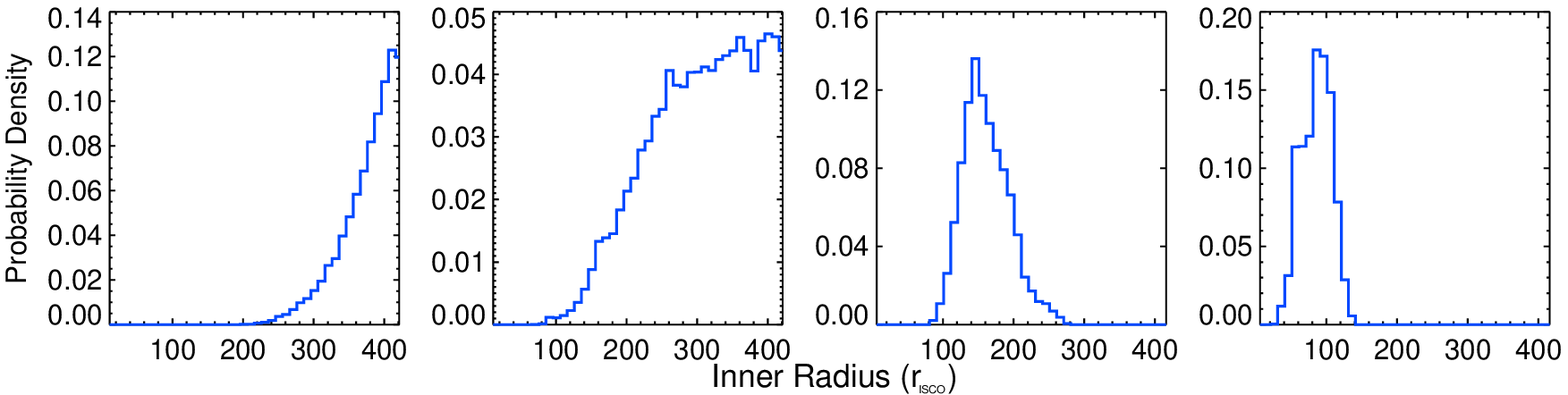, width=0.95\textwidth}}
\caption{Inner radius results from our MCMC based analysis displayed from left to right for Observations 1 to 4 respectively. Top panels: The distribution of inner radius values along the 200,000 element chain with every 200th element displayed for clarity. Lower panels: Probability distributions for the inner radius, each bin has a width of $10\,r_{\rm isco}$. An inner accretion disc at the ISCO is clearly ruled out for the four hard state observations.}
\label{Fig:mcmc1}
\end{figure*}
\begin{figure*}
\centering
{\epsfig{file=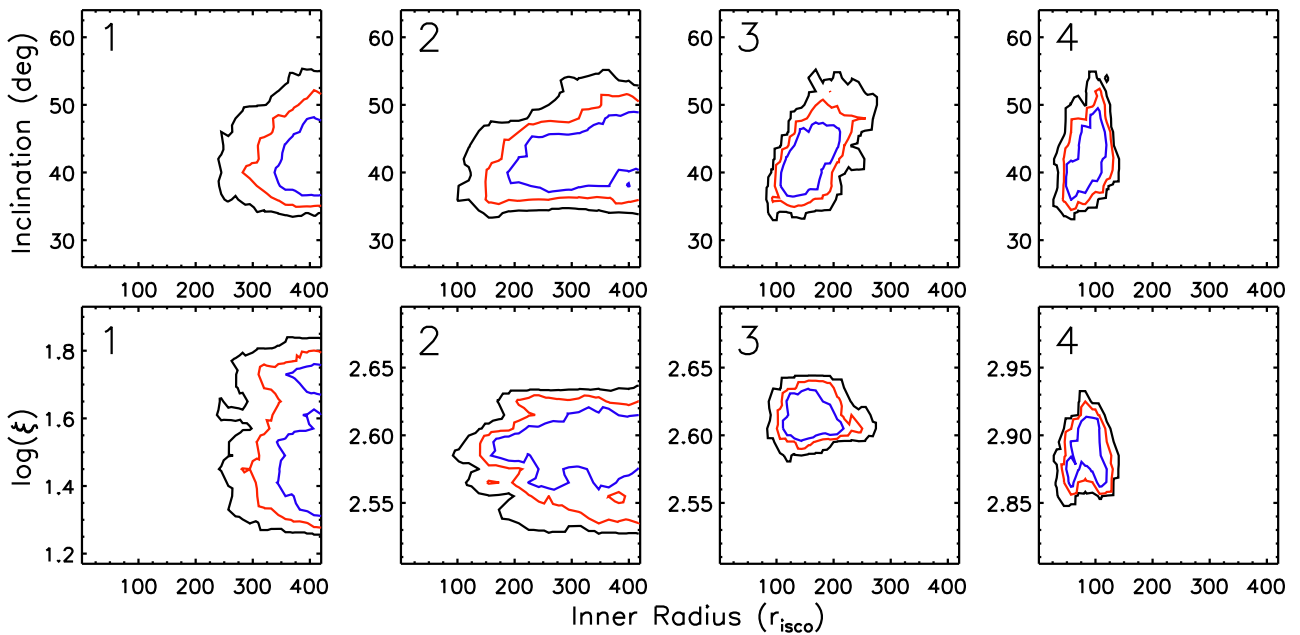, width=0.95\textwidth}}
\caption{Two-dimensional probability distributions of the inclination (top) and ionisation (bottom) parameters with the inner radius. For each panel we plot contour levels at the 68\% (blue), 90\% (red) and 99\% (black) confidence levels. Each plot contains the full distribution for the whole parameter range with the axis reduced for clarity. In all eight distributions there is no indication of a significant correlation between the parameters.}
\label{Fig:mcmc2}
\end{figure*}

Figure\,\ref{Fig:mcmc1} (top panels) displays the values of the inner radius for the four observations from every 200th step of the 200,000 elements. Each distribution shows no sign of significant deviation or trend, thus suggesting that the global minimum has in fact been found, offering even more certainty to the inner radius evolution we have uncovered. The lower panels of Fig.\,\ref{Fig:mcmc1} show the probability distribution for the inner radius, in-keeping with Table \ref{fits}. For each observation the inner disc being at the ISCO is clearly ruled out.

Another benefit of MCMC analysis is that it allows one to reveal any correlations between parameters. In Fig.\,\ref{Fig:mcmc2} we present two-dimensional probability distributions of the inner radius with the inclination and ionisation parameters. Each axis is chosen for clarity, rather than to span the entire parameter space of the model, but still encompasses the full probability distribution for both parameters. Each distribution shows little sign of a correlation, indicating that no significant parameter degeneracies are at play, and that we have indeed found the global best fit.


\subsection{Does disc evolution monotonically track luminosity? - Comparing rise and decay}\label{rise_decay}

\begin{figure}
\centering
\centerline{\epsfig{file=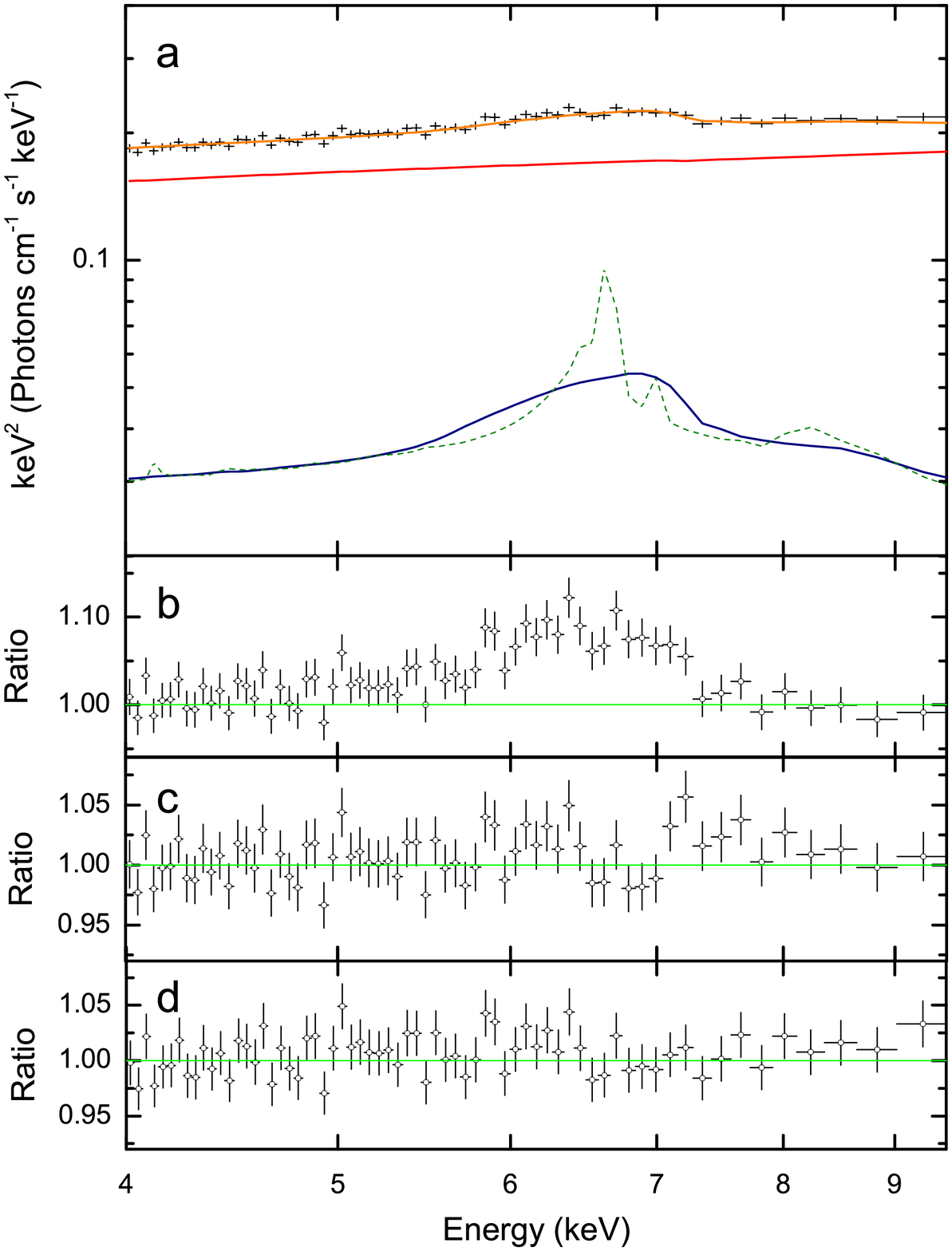, width=0.45\textwidth}}
\caption{(a) The unfolded spectral fit for the 2011 HIMS observation (Observation 2b) using \textsc{relconv$\ast$xillver} to model the reflection component. The colour scheme follows that of Fig.\,\ref{Fig:EEUF}. (b) The data/model ratio for the continuum fit in Table \ref{table_continuum} displaying strongly broadened Fe K$\alpha$ emission. (c) The model residuals when using \textsc{relline} to model the Fe line. (d) The data/model ratio when the reflection is modelled using \textsc{relconv$\ast$xillver}. The spectra have been re-binned for the purposes of plotting.}
\label{Fig:HIMS}
\end{figure}
So far we have only analysed the four hard state observations, ignoring the 2011 hard-intermediate dataset (Observation 2b). This observation took place in decay at a very comparable luminosity to Observation 2, which was taken whilst GX 339-4 was in its rise through the hard state. Therefore, if disc evolution is solely a trend with luminosity then one should expect to obtain similar results between the two observations. Panel (b) of Fig.\,\ref{Fig:HIMS} shows the data/model ratio of the continuum fit to Observation 2b. We find a strongly broadened Fe line, much more asymmetric than the hard state profiles, indicating relativistic broadening is more significant. We fix the inclination to that found in Table \ref{fits} to probe this conclusion, and the results are presented in Table \ref{2009-2011}. The same fixed values are used for the spin (0.9), emissivity index (3) and disc outer radius ($1000\,r_{\rm g}$).

We find that for both the line and reflection fitting techniques the inner accretion disc radius is estimated to be significantly smaller for the intermediate state observation. This strong contrast in the inner radius of the disc suggests a possible hysteresis in how the inner disc evolves throughout the outburst. Furthermore, the disc inner radius found through both methods is smaller than for all of the hard state observations at 90\% confidence. Therefore, this strongly suggests that the disc is truncated throughout the hard state, even in Observation 4, which is at $\sim$$0.15$ L$_{\rm Edd}$ (assuming a BH mass and distance of $8$\,$M_{\sun}$ and $8$\,kpc respectively). This conclusion is further supported by the small difference in fit between the two methods (only $\Delta\chi^2=+7$ for no additional degrees of freedom using {\sc relline}; Table \ref{2009-2011}), meaning that the profile is now dominated by relativistic effects from emission close to the BH.

The ionisation parameter is similar ($\sim$log(2.6)), although slightly larger in decay. The decay spectrum is softer (as defined by the hard-intermediate state) since there is more thermal emission from the disc (compare {\it e.g.} \citealt{Hiemstra11} with \citealt{Reis10}). This will lead to increased heating of the surface layers of the disc where the reflection spectrum originates from, and therefore display broader line profiles \citep{Ross07}, similar to that of a larger ionisation parameter. The current publicly available models do not take this into account (see $\S$\ref{discussion} for further discussion of this), and it is likely that the hard-intermediate state observation would then require a larger ionisation parameter to solve for this.

\begin{table}
\centering
\caption{Results from fitting Observation 2b with the \textsc{relline} (left section) and \textsc{relconv$\ast$xillver} (right section) models respectively.}
\begin{tabular}{lccc}\hline\hline
Parameter					& \textsc{relline}			& \textsc{relconv$\ast$xillver}		\\\hline	
$\Gamma$					& 1.84$\pm{0.02}$		& 1.85$^{+0.04}_{-0.03}$			\\
N$_{\rm PL}$					& 0.15$\pm{0.01}$		& 0.13$\pm{0.03}$			 	\\
E (keV)						& 6.97$_{-0.06}$			& ...		 						\\
$r_{\rm in}$ ($r_{\rm isco}$)	& 2.3$^{+0.3}_{-0.4}$		& 12.5$^{+7.2}_{-5.8}$		 	\\
N$_{\rm L}$ ($10^{-4}$)		& 6.69$^{+1.05}_{-1.00}$	& ...		 						\\
$\log(\xi)$					& ...						& 2.65$^{+0.22}_{-0.52}$			\\
N$_{\rm R}$ ($10^{-6}$)		& ...						& 1.50$^{+5.01}_{-0.50}$		 	\\
E.W. (eV)					& 152$^{+29}_{-30}$		& ...		 						\\
RF							& ...						& 0.23$\pm{0.12}$	 			\\\hline
$\chi{^2}/\nu$				& 1563/1518				& 1556/1518		 				\\\hline
\end{tabular}
\tablefoot{The inclination was frozen to that fitted simultaneously by the four hard state observations for the respective model (see Table \ref{fits}). The emissivity and disc outer radius were frozen to be $r^{-3}$ and 1000\,$r_{\rm g}$ respectively, following the previous sections.}
\label{2009-2011}
\end{table}


\subsection{Power density spectra}\label{timing_sec}
The power spectra of BHs in the hard state display similar characteristics \citep{VDK95,VDK06}. Typically, they are described by a flat region of power $P(\nu) \propto \nu^0$  below a frequency $\nu_{\rm b}$, known as the `break frequency'. Above $\nu_{\rm b}$ the power spectra steepens to roughly $P(\nu) \propto \nu^{-1}$. This break frequency is coupled with the BH mass and accretion rate \citep{McHardy06,Koerding07}, and gradually increases as the source progresses through the hard state (see {\it e.g.} \citealt{Migliari05}). Furthermore, a number of correlations have been found, notably with a steepening of the Comptonised spectrum and an increasing amount of reflection \citep{Gilfanov99}. Both of these trends can be interpreted as an increasing penetration of the cool accretion disc into the inner hot flow, which provides more seed photons to cool the flow. Therefore, it has been suggested that $\nu_{\rm b}$ is associated with the truncation radius of the inner disc \citep{Gilfanov99,Churazov01}.

\begin{figure}
\centering
\centerline{\epsfig{file=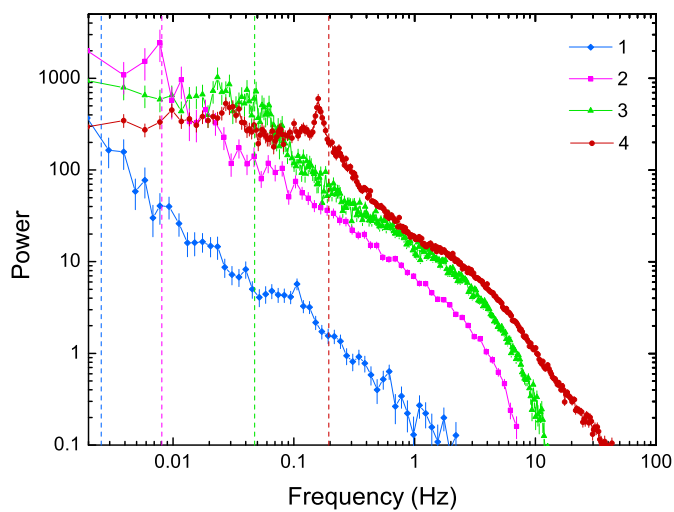, width=0.5\textwidth}}
\caption{The PDS of the four RXTE observations in Table \ref{timing}, clearly showing that the break frequency shifts to higher frequencies in the higher luminosity observations. The dashed lines indicate the fitted break frequency for each observation (Table \ref{timing}). The same colour/shape scheme as Fig.\,\ref{Fig:HID} is used.}
\label{Fig:PDS}
\end{figure}

For each of the hard state observations listed in Table \ref{observations} we analyse the power density spectra (PDS) of a simultaneous RXTE observation. Where more than one observation is available we chose the one with the best signal-to-noise. Unfortunately, the nearest observation to Observation 2b was not of sufficient quality to be used, and hence this epoch is ignored. Figure\,\ref{Fig:PDS} displays all four PDS and a clear evolution to higher frequencies can be seen for $\nu_{\rm b}$ as the luminosity increases. PDS fitting was performed via standard $\chi^2$ fitting with XSPEC. We fitted the noise components with one zero-centred Lorentzian plus other two centred at $\sim$ few Hz (see \citealt{Belloni02} for details on the standard procedure). For Observation 4 another narrow Lorentzian was added due to the presence of a weak quasi periodic oscillation. The break fitted for each PDS follows the expected trend to higher frequencies with luminosity (Table \ref{timing}).

\begin{table}
\centering
\caption{Fitted break frequencies ($\nu_{\rm b}$) for RXTE observations simultaneous to that of the high-resolution spectra listed in Table \ref{observations}.}
\begin{tabular}{ccccc}\hline\hline
				& ObsID			& Date							& $\nu_{\rm b}$ (Hz)									\\\hline	
\multirow{4}{*}{1}	& 93702-04-01-00	& 2008-09-24 00:29:53		& \multirow{4}{*}{0.00255$_{-0.00048}^{+0.00065}$}		\\
				& 93702-04-01-01	& 2008-09-25 06:23:06		&													\\
				& 93702-04-01-02	& 2008-09-25 11:06:06		&													\\
				& 93702-04-01-03	& 2008-09-25 19:00:14		&													\\
2				& 94405-01-03-00	& 2009-03-26 11:04:44		& 0.00808$_{-0.0017}^{+0.0026}$						\\
3				& 90118-01-07-00	& 2004-03-18 18:01:16		& 0.0474$^{+0.0028}_{-0.0026}$						\\
4				& 95409-01-12-01	& 2010-03-28 03:13:35		& 0.195$_{-0.006}^{+0.007}$							\\\hline
\end{tabular}
\tablefoot{In order to analyse the lowest flux epoch we combined four contemporaneous observations. Quoted errors are at the $1\sigma$ level.}
\label{timing}
\end{table}

\begin{figure}
\centering
\centerline{\epsfig{file=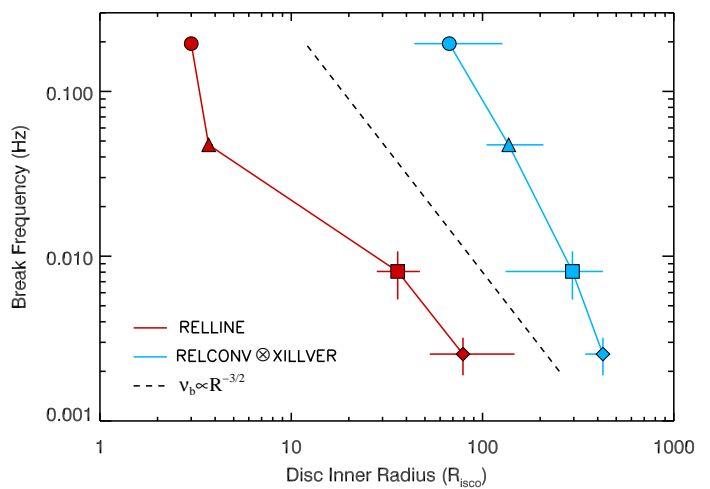, width=0.5\textwidth}}
\caption{The estimated disc inner radii (x-axis; Table \ref{fits}) using \textsc{relline} (red) and {\sc relconv$\ast$xillver} (green) versus and the fitted break frequency from simultaneous RXTE observations (y-axis; Table \ref{timing}). The dotted line represents the relation $\nu \propto $ R$^{-3/2}$, which corresponds to the dynamical and viscous timescales for accretion onto a BH.}
\label{Fig:rin_break}
\end{figure}

Earlier in this work we have used the reflection spectrum to examine how the inner radius of the disc evolves through the hard state. If $\nu_{\rm b}$ corresponds to the inner radius of the disc it allows us an independent measure of this, and hence in Fig.\,\ref{Fig:rin_break} the two are plotted to see how they correlate. In particular, the trend between the break frequency and the inner radius from \textsc{relconv$\ast$xillver} is very strong, and suggests the two are indeed connected. The dynamical and viscous timescales (both of which could be tracking the inner radius) for accretion onto a BH are proportional to $R^{3/2}$. We also plot this in Fig.\,\ref{Fig:rin_break}, which indicates that the slope of the break and inner radius is slightly steeper. As we described before, other parameters have been found to correlate with the break frequency, and thus it is not certain which is truly causal in driving the change in frequency of the break. However, the softening power-law photon index and increasing reflection fraction trends that we mention before are both expected if the inner radius of the disc decreases \citep{Done07}. Mass accretion rate may also influence the break frequency, or be intrinsically linked to the inner radius \citep{Migliari05,McHardy06}.

Nevertheless, the evolution is remarkably similar. To see how the trend continues we fit the final hard state RXTE observation of the 2010 outburst (ObsID: 95409-01-13-06), which is the same outburst as Observation 4. The break frequency is now found to be at $0.53\pm0.03$, more than a factor of two larger than Observation 4, clearly indicating that the evolution continues through the hard state. Therefore, if the inner radius of the accretion disc and the break frequency are physically linked, the inner disc radius must continue to evolve throughout the hard state. Furthermore, as the source transitions into the intermediate states the break frequency is seen to continue above 1 Hz (see also \citealt{Belloni05,VDK06}). 


\subsection{Testing the full bandpass}\label{full_test}
So far we have restricted our analysis to above 4\,keV, since this allows us to fit a simpler continuum model in the computationally intensive task of fitting four datasets with tied parameters. Including the data below 4\,keV firstly adds a considerable amount of extra degrees of freedom that need to be fit. But more importantly it presents a much more complex continuum. Above 4\,keV the continuum is a simple power-law, but below this the thermal emission from the disc quickly becomes significant, and adds considerably more curvature to the continuum. Furthermore, the requirement of a \textsc{diskbb} model to fit this, and the much stronger effect of interstellar absorption, requires three additional free parameters to be fit per spectrum. All of these factors combined make simultaneous fitting of the four hard state spectra, with tied parameters and a self-consistent treatment for reflection (as applied in \S\ref{full_ref}), impossible in a reasonable timescale if the full bandpass is used.

Another potential issue is that the weak thermal component can be mistaken for the soft excess in the reflection model. Also the lower part of the bandpass contains strong edges in the effective area and is hence generally less well calibrated. Since the statistics are better at lower energies these features could drive the reflection fit rather than the more revealing Fe K emission. Whilst ignoring the data below 4 keV restricted our bandwidth, it significantly decreased our calibration dependance as well. This procedure is standard practice in the AGN community where the majority of X-ray reflection studies have taken place, be it in systematic investigations (see {\it e.g.} \citealt{Nandra07,Calle10}) or single observations (see {\it e.g.} \citealt{Fabian02,Ponti09}).

The focus of this study is to systematically uncover how the inner disc is evolving in one source, and hence tying parameters and minimising calibration and model dependance is of utmost importance. As we have outlined, by necessity this requires fitting above 4\,keV, but this then also throws away a large amount of the soft X-ray spectrum. Whilst the reflection spectrum is weaker than the thermal and Comptonised emission below 4\,keV, it is still significant, and in particular this part of the reflection spectrum holds important information about the ionisation stage. Furthermore, the most broadened Fe profiles, i.e. that of a maximally rotating black hole, may be mistaken for the continuum when only fitted above 4\,keV. It is therefore essential that be examine the full bandpass of each spectrum to ensure that our results from fits above 4\,keV are not in error.  

To examine this consistency we thus fit each spectrum individually using a full bandpass, as is the standard approach in snapshot X-ray studies of BHXRBs. For the \emph{XMM-Netwon} observations the full bandpass covers the 1.3--1.75 and 2.35--10\,keV ranges. The \emph{Suzaku} observations were fitted in the 0.7--1.7 and 2.4--10\,keV bands (see \S\ref{obs_data} for more details on the bandpass used). Since the inclination cannot be jointly determined we fix it to the value determined with \textsc{xillver} (Table \ref{fits}; 42$^{\circ}$) to allow a fair systematic comparison of each dataset. Because the inclination is measured from the sharper blue wing of the Fe K$\alpha$ profile this should be well determined even if the analysis above 4\,keV is in error. We nevertheless test the effect of different inclination angles over the full bandpass in \S\ref{inclination}.

\begin{figure}
\centering
\centerline{\epsfig{file=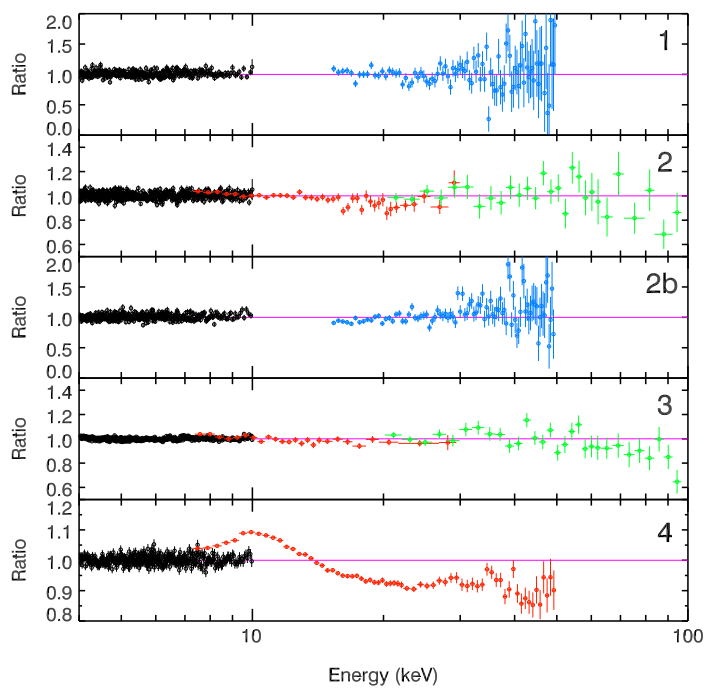, width=0.5\textwidth}}
\caption{Data/model ratio plots of the PCA/HEXTE/PIN data added to the best-fit model fitted between 4--10\,keV with the \emph{XMM-Newton} and \emph{Suzaku} data (Tables\,\ref{fits} and \ref{2009-2011}). The PCA/HEXTE/PIN data are allowed a free constant, but all other parameters are frozen to the values listed in (Tables\,\ref{fits} and \ref{2009-2011}). Observations 1--3 show very good agreement with the hard X-ray data, suggesting that the model fit between 4--10\,keV is correct. However, in Observation 4 there appears to be a large discrepancy between the pn and PCA data. Hard X-ray data are colour-coded as PCA (red), HEXTE (green) and PIN (blue), and fit in the energy ranges outlined in \S\ref{obs_data}.}
\label{Fig:gx339_HE}
\end{figure}

With the added disc component our full model is now \textsc{phabs(diskbb+powerlaw+relconv$\ast$xillver)} which we fit with the additional parameters for the inner disc temperature and normalisation allowed to be free. The column density is now also fitted freely due to the strong impact of interstellar absorption below 4\,keV, and the results can be found in Table \ref{full_band}. Unfortunately, the weak disc component cannot be constrained in Observations 2 and 3 due to the lower cut-off of 1.3\,keV used for the pn data. The remaining observations are all well fitted by the added disc model, all reporting a reduced-$\chi^2$ of $\leq1.10$, except for Observation 3. The main cause of the poor fit to Observation 3 is in the soft region of the bandpass ($\textless$ 4\,keV), and is probably due to some un-modelled thermal emission being present there. The Fe line is still well modelled. The column density (N$_{\rm H}$) varies too, which is expected to remain fairly constant \citep{Miller09b}. We expect this is due to a combination of factors. The lower-bound of 1.3\,keV restricts the ability to constrain the N$_{\rm H}$ in the \emph{XMM-Newton} spectra. Also, where it has not been possible to fit a disc component (Observations 2 and 3), the N$_{\rm H}$ is likely to respond to any soft X-ray excess remaining. This may also be a response to calibration issues in the timing mode. We note that \cite{Kolehmainen13} use some of these datasets, and apply more complex continuum models, but still find strong evidence for a truncated disc through the Fe line.

\renewcommand{\arraystretch}{1.5}
\begin{table*}
\centering
\caption{Results from fitting the four hard state observations individually with \textsc{phabs(diskbb+powerlaw+relconv$\ast$xillver)} using a full bandpass. Results are presented for emissivity profiles of $r^{-3}$ and $r^{-2.5}$ (bottom).}
\begin{tabular}{cc ccccc}\hline\hline
Model				& Parameter							& 1						& 2						& 3					& 4						& 2b						\\\hline\hline
\multicolumn{7}{c}{$q=3$} \\ \hline
\textsc{phabs}			& N$_{\rm H}$ ($\times10^{22}$)	& 0.62$\pm{0.04}$		& 0.30$\pm{0.01}$		& 0.24$\pm{0.01}$	& 0.32$\pm{0.03}$		& 0.38$^{+0.02}_{-0.01}$		\\
\textsc{diskbb}			& $T_{\rm in}$ (keV)				& 0.16$\pm{0.01}$		& ...						& ...					& 0.22$^{+0.02}_{-0.04}$	& 0.27$\pm{0.01}$			\\
					& N$_{\rm BB}$ ($\times10^3$)			& 3.72$^{+2.12}_{-1.52}$	& ...						& ...					& 14.0$^{+27.0}_{-7.5}$	& 3.94$^{+1.50}_{-0.96}$		\\
\textsc{powerlaw}		& $\Gamma$							& 1.68$\pm{0.02}$		& 1.45$\pm{0.01}$		& 1.49$\pm{0.01}$	& 1.60$\pm{0.01}$		& 1.90$\pm{0.01}$			\\
					& N$_{\rm PL}$						& 0.013$\pm{0.001}$		& 0.086$\pm{0.003}$		& 0.16$\pm{0.01}$	& 0.41$\pm{0.07}$		& 0.13$\pm{0.01}$			\\
\textsc{relconv}			& $\theta$ ($^{\circ}$)				& \multicolumn{5}{c}{42}																									\\
					& $r_{\rm in}$ ($r_{\rm isco}$)			& \textgreater 320			& 213$^{+218}_{-99}$		& 110$^{+30}_{-15}$	& 69$^{+37}_{-29}$		& 17$^{+9}_{-5}$				\\
\textsc{xillver}			& $\log(\xi)$						& 1.83$^{+0.04}_{-0.10}$	& 2.61$^{+0.03}_{-0.01}$	& 2.63$\pm{0.01}$	& 2.86$\pm{0.03}$		& 2.78$^{+0.04}_{-0.10}$		\\
					& N$_{\rm R}$ ($10^{-6}$)				& 1.60$^{+0.57}_{-0.29}$	& 3.29$^{+0.22}_{-0.24}$	& 5.11$\pm{0.08}$	& 20.3$^{+0.9}_{-0.8}$	& 1.33$^{+0.33}_{-0.19}$		\\ \hline
\multicolumn{2}{c}{$\chi{^2}/\nu$}							& 2275/2227				& 1774/1614				& 2920/1614			& 1671/1612				& 2378/2226					\\\hline\hline
\multicolumn{7}{c}{$q=2.5$} \\ \hline
\textsc{phabs}			& N$_{\rm H}$ ($\times10^{22}$)	& 0.62$\pm{0.04}$		& 0.30$\pm{0.01}$		& 0.24$\pm{0.01}$	& 0.32$\pm{0.03}$		& 0.38$^{+0.02}_{-0.01}$		\\
\textsc{diskbb}			& $T_{\rm in}$ (keV)				& 0.16$\pm{0.01}$		& ...						& ...					& 0.22$\pm{+0.02}$		& 0.27$\pm{0.01}$			\\
					& N$_{\rm BB}$ ($\times10^3$)			& 3.72$^{+1.95}_{-1.32}$	& ...						& ...					& 14.0$^{+41.5}_{-7.4}$	& 3.81$^{+1.55}_{-0.89}$		\\
\textsc{powerlaw}		& $\Gamma$							& 1.68$\pm{0.02}$		& 1.45$\pm{0.01}$		& 1.49$\pm{0.01}$	& 1.60$^{+0.01}_{-0.02}$	& 1.89$\pm{0.01}$			\\
					& N$_{\rm PL}$						& 0.013$\pm{0.001}$		& 0.087$\pm{0.003}$		& 0.16$\pm{0.01}$	& 0.41$^{+0.10}_{-0.07}$	& 0.13$\pm{0.01}$			\\
\textsc{relconv}			& $\theta$ ($^{\circ}$)				& \multicolumn{5}{c}{42}																									\\
					& $r_{\rm in}$ ($r_{\rm isco}$)			& \textgreater 317			& 211$^{+215}_{-104}$	& 107$^{+27}_{-19}$	& 53$^{+32}_{-46}$		& 14$^{+7}_{-6}$				\\
\textsc{xillver}			& $\log(\xi)$						& 1.83$^{+0.04}_{-0.05}$	& 2.60$^{+0.04}_{-0.03}$	& 2.63$\pm{0.01}$	& 2.86$^{+0.03}_{-0.05}$	& 2.80$^{+0.03}_{-0.12}$		\\
					& N$_{\rm R}$ ($10^{-6}$)				& 1.60$^{+0.30}_{-0.26}$	& 3.29$^{+0.23}_{-0.24}$	& 5.11$\pm{0.08}$	& 20.4$\pm{0.8}$			& 1.25$^{+0.41}_{-0.12}$		\\ \hline
\multicolumn{2}{c}{$\chi{^2}/\nu$}							& 2276/2227				& 1774/1614				& 2921/1614			& 1671/1612				& 2380/2226					\\ \hline
\end{tabular}
\tablefoot{The photon index in the reflection models is tied to that of the continuum power-law and the Fe abundance is assumed to be solar. The same fixed values are used as for Table \ref{fits}. In some cases the inner radius parameter has reached the largest tabulated value in the model ($1000\,r_{\rm g}$), therefore we only present the lower limit in this case.}
\label{full_band}
\end{table*}

For all of the observations, we find that each parameter remains consistent at the 90\,\% confidence level with those in \S\ref{full_ref}, and clearly indicates that our results in that section were not driven by fitting only above 4 keV. In particular, the inner radius parameter shows little change, and the trend of the disc extending closer to the BH with increasing luminosity still remains. Additionally, the power-law photon index shows very little variation, indicating that the continuum was still very well constrained when we restricted the fit to above 4 keV. Furthermore, if a broad red wing from a disc at the ISCO was present, then it is possible that it could have been masked by a harder photon index when fitting only above 4 keV. However, the extremely consistent results found when fitting the full bandpass verifies that the profiles in the hard state show little relativistic broadening, confirming that the disc is truncated throughout the hard state.

Recently, some hard-state observations with the broad bandpass of \emph{Suzaku} have found that a single power-law does not sufficiently fit the data, and instead a double thermal Comptonsiation model (with different optical depths) is required to fit the soft and hard continua. In \S\ref{Sec:Fe_fits} we employed a soft bandpass between 4--10\,keV to fit the data, thus our analysis is unlikely to resolve more a complex continuum such as this. Instead, we add the hard X-ray data from \S\ref{continuum} to the best-fit model determined over 4--10\,keV (Tables\,\ref{fits} and \ref{2009-2011}) to see how the hard X-ray data match up. We allowed a free constant between the data, but otherwise all the parameters were fixed to the values determined between 4--10\,keV in \S\ref{Sec:Fe_fits}. In Fig. \ref{Fig:gx339_HE} we display the data/model residuals, which show that Observations 1--3 correctly fit the data above 10\,keV, and suggests that the continuum power-law model that we employed is correct; however, Observation 4 shows a significant discrepancy above 10\,keV. Interestingly, even below 10\,keV the PCA and pn data do not agree which would suggest that the two datasets are inconsistent, and, therefore, it is not possible to determine if the model is correct. Each PCA/pn constant is 1.12, 1.10 and 1.16 for Observations 2, 3 and 4, respectively. The PCA and pn data have previously been noted to be discrepant \citep{Hiemstra11,Kolehmainen13}, and this may be the source of this inconsistency, although this does not explain why it is only seen in Observation 4.

\subsubsection{The unknown inclination of GX 339-4}\label{inclination}

The orbital inclination of GX 339-4 is unknown, thus any definitive constraint on the accretion disc inclination is missing. However, indirect evidence suggests intermediate values. On one hand, the absence of absorption dips, eclipses and absorption features from an accretion disc, together with the observed colours and outburst evolution, indicate that the inclination is less than $\sim70^{\circ}$ \citep{Ponti12,Munoz13}. On the other hand, an inclination lower than $\sim40^{\circ}$ is unlikely since it would result in a very large BH mass \citep{Hynes03,Munoz08}. The respective strengths of the different relativistic effects is determined by the inclination of the accretion disc, one obvious example being aberration. Each effect ultimately transforms, and most importantly broadens, the line profile in different ways, thus the inclination parameter is very degenerate with the inner radius. A key aspect of this work is that we jointly determine the inclination across all four hard state spectra (see \S\ref{full_ref}); however, this only feasible between $4-10$\,keV. We thus instead test the effect of the inclination by fitting each full bandpass dataset individually for a range of fixed inclination values, investigating the full range of $\cos{i}$ in steps of 0.05.

\begin{figure}
\centering
\centerline{\epsfig{file=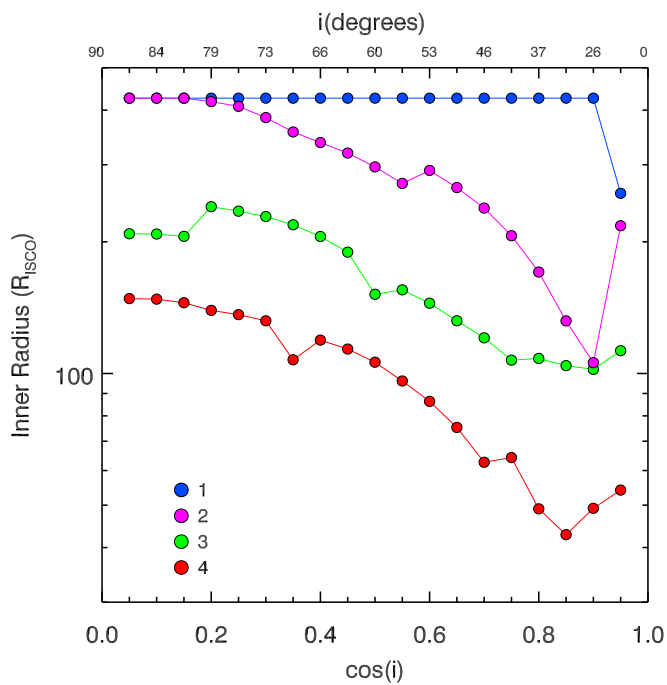, width=0.5\textwidth}}
\caption{The fitted inner radius parameter for different fixed values of inclination, fitted individually using a full bandpass. The coloured symbols indicate the four observations in Table \ref{observations}: blue (1), magenta (2), green (3) and red (4). For clarity errors are not plotted but are instead listed in Table \ref{table_cosi}.}
\label{Fig:rin_cosi}
\end{figure}
\renewcommand{\arraystretch}{1.3}
\begin{table}
\centering
\caption{The fitted inner radius of each observation (Table \ref{observations}) for different fixed values of inclination.}
\begin{tabular}{ccccc}\hline\hline
$\cos(i)$	&	1									&	2									&	3							&	4							\\\hline
0.05		&				$	\textgreater	373	$	&				$	\textgreater	328	$	&	209	$^{+	49	}	_{-	35	}$	&	148	$^{+	136	}	_{-	81	}$	\\
0.1		&				$	\textgreater	377	$	&				$	\textgreater	327	$	&	208	$^{+	52	}	_{-	32	}$	&	148	$^{+	131	}	_{-	73	}$	\\
0.15		&				$	\textgreater	382	$	&				$	\textgreater	341	$	&	206	$^{+	53	}	_{-	30	}$	&	145	$^{+	126	}	_{-	66	}$	\\
0.2		&				$	\textgreater	380	$	&	418	$^{+	8	}	_{-	94	}$			&	241	$^{+	28	}	_{-	70	}$	&	139	$^{+	126	}	_{-	60	}$	\\
0.25		&				$	\textgreater	376	$	&	408	$^{+	18	}	_{-	105	}$			&	235	$^{+	28	}	_{-	56	}$	&	136	$^{+	121	}	_{-	59	}$	\\
0.3		&				$	\textgreater	373	$	&	385	$^{+	41	}	_{-	108	}$			&	228	$^{+	27	}	_{-	54	}$	&	132	$^{+	116	}	_{-	57	}$	\\
0.35		&				$	\textgreater	371	$	&	357	$^{+	69	}	_{-	102	}$			&	219	$^{+	27	}	_{-	52	}$	&	107	$^{+	130	}	_{-	37	}$	\\
0.4		&				$	\textgreater	374	$	&	337	$^{+	89	}	_{-	102	}$			&	206	$^{+	28	}	_{-	49	}$	&	119	$^{+	102	}	_{-	50	}$	\\
0.45		&				$	\textgreater	368	$	&	319	$^{+	107	}	_{-	102	}$			&	189	$^{+	30	}	_{-	43	}$	&	114	$^{+	96	}	_{-	49	}$	\\
0.5		&				$	\textgreater	364	$	&	297	$^{+	129	}	_{-	98	}$			&	152	$^{+	44	}	_{-	18	}$	&	106	$^{+	88	}	_{-	45	}$	\\
0.55		&				$	\textgreater	362	$	&	272	$^{+	154	}	_{-	91	}$			&	155	$^{+	35	}	_{-	27	}$	&	96	$^{+	82	}	_{-	39	}$	\\
0.6		&				$	\textgreater	354	$	&	291	$^{+	135	}	_{-	127	}$			&	145	$^{+	30	}	_{-	25	}$	&	86	$^{+	73	}	_{-	34	}$	\\
0.65		&				$	\textgreater	345	$	&	266	$^{+	160	}	_{-	146	}$			&	132	$^{+	27	}	_{-	20	}$	&	75	$^{+	61	}	_{-	28	}$	\\
0.7		&				$	\textgreater	337	$	&	239	$^{+	187	}	_{-	108	}$			&	121	$^{+	25	}	_{-	17	}$	&	63	$^{+	57	}	_{-	20	}$	\\
0.75		&				$	\textgreater	318	$	&	207	$^{+	219	}	_{-	94	}$			&	107	$^{+	30	}	_{-	14	}$	&	64	$^{+	38	}	_{-	25	}$	\\
0.8		&				$	\textgreater	291	$	&	171	$^{+	240	}	_{-	76	}$			&	108	$^{+	13	}	_{-	18	}$	&	49	$^{+	38	}	_{-	15	}$	\\
0.85		&				$	\textgreater	241	$	&	132	$^{+	189	}	_{-	56	}$			&	104	$^{+	18	}	_{-	22	}$	&	43	$^{+	33	}	_{-	10	}$	\\
0.9		&				$	\textgreater	165	$	&	106	$^{+	146	}	_{-	45	}$			&	102	$^{+	14	}	_{-	15	}$	&	49	$^{+	20	}	_{-	15	}$	\\
0.95		&	258	$^{+	168	}	_{-	149	}$			&	218	$^{+	93	}	_{-	142	}$			&	113	$^{+	18	}	_{-	29	}$	&	54	$^{+	34	}	_{-	16	}$	\\\hline
\end{tabular}
\tablefoot{In some cases the inner radius parameter has reached the largest tabulated value in the model ($1000\,r_{\rm g}$), therefore we only present the lower limit in this case.}
\label{table_cosi}
\end{table}

Figure\,\ref{Fig:rin_cosi} displays how the absolute value of the inner radius varies throughout the range of $\cos{i}$. The influence of the inclination parameter is immediately clear, fitting smaller inner radii as the inclination decreases. As previously described in \S\ref{full_ref}, this is a consequence of the reduced doppler shift when a disc is face-on, thus as the inclination increases (becoming more edge-on) a larger inner radius is required to replicate the observed profile width. This degeneracy underlines the importance of the simultaneous and systematic approach we have taken in \S\ref{full_ref}, and highlights the limitation of snap-shot analysis often habitual in X-ray spectroscopy. In Table \ref{table_cosi} we detail the complete results of the analysis. We note that although Observation 1 appears to remain at a constant inner radius regardless of inclination, this is because the parameter is pegged at the hard limit. The consistent relationship between the inner radius and inclination means that the inner radius ubiquitously decreases with increasing source luminosity for any assumed disc inclination. Furthermore, this again confirms that the trend presented in \S\ref{full_ref} and Fig.\,\ref{Fig:rin_lum} is not an artefact of the reduced bandpass.

\subsubsection{The emissivity parameter}\label{emissivity}

Throughout this study we have fixed the emissivity parameter of each observation due to the high level of degeneracy it has with the inner radius parameter. Additionally, previous works fitting the Fe line in the hard state have regularly found emissivity parameter values consistent to the $r^{\rm-3}$ law that we have adopted \citep{Miller06,Reis09,Reis10,Done10,Shidatsu11}. A larger emissivity index yields increased emission from the inner regions of the disc, and hence increases the significance of the relativistic effects upon the overall profile. This is somewhat analogous to having an inner disc radius closer to the BH, where the influence of relativity becomes stronger. We do acknowledge, however, that the profile we have assumed may be incorrect, and could evolve throughout the outburst depending on the geometry of the hard X-ray source.

The value of $3$ that we use for the emissivity parameter is the value for a flat Newtonian disc being irradiated by a point source. In General Relativity, however, one would expect the hard X-ray source to be more focused on the inner regions because of light-bending, resulting in a steeper fall off with radius (\ie a larger emissivity index; \citealt{Miniutti03}). Furthermore, decreasing the scale height of the corona would further this effect. To account for this, a broken power-law profile may instead be required to describe the contrasting emission from the inner region. A recent studies by \cite{Dauser10} and \cite{Wilkins12} investigate further the emissivity profiles for different coronal geometries, the general trend being towards a value of roughly $3$ outwards of $\sim10-30\,r_{\rm g}$. Such radii are much smaller than the disc inner radius we find in this study. Additionally, as discussed in \cite{Fabian12}, it is only in the innermost region ($\textless 2r_{\rm g}$) where a steeper inner profile is strictly necessary. Nevertheless, it is possible that a varying emissivity profile may be contributing to the trend we observe.

To test how the emissivity profile may affect our results, we re-analysed the full-bandpass observations, replacing the $r^{-3}$ law with $r^{-5/2}$ (Table \ref{full_band}). The fitted radii were largely unchanged, as were the remaining free parameters, but generally moved to smaller values, although well within the errors of the $r^{-3}$ fit. The change is a consequence of the flatter profile giving a weaker weighting to the emission from smaller radii, thus to model the broadening of the profile the inner radius parameter reduces. Observations 2b and 4 saw a larger change in radii, which is due to the stronger effect upon the modelled profile by changing the emissivity index, because the integrated disc region is larger (i.e. the range $r_{\rm in} - r_{\rm out}$ is larger). The fit itself is equally good, and thus it doesn't seem that there is any significant effect to our results in adopting a flatter emissivity index.

To investigate the emissivity profile further we fit a free once-broken emissivity profile to each observation, again using a full bandpass. The emissivity indexes, both represented by the parameter $q$ in the form $r^{-q}$, are fitted in the ranges $3-10$ and $2-3.5$ for the inner and outer region respectively. The larger range selected for the inner index reflects the influence of light-bending, and in any case the index should be greater than $2$ for the outer region of the disc, but not significantly larger than the Euclidean value of 3 \citep{Wilkins12}. The break radius $r_{\rm br}$ is allowed to span the range $2.32-10\,r_{\rm g}$, whereby the lower bound is the ISCO for a spin of $0.9$. For all four observations the inner region index and break radius are unable to be resolved, both being insensitive to the fit as a significantly truncated inner radius, much larger than $r_{\rm br}$, is still preferred. To stabilise the fit we fixed the inner index to the smallest expected value of $3$ and break radius to $6r_{\rm g}$, fitting the outer index as previously. We note that the choice of the inner index and break radius values do not affect the results since $r_{\rm in}\textgreater r_{\rm br}$.

The inner radius of Observation 1 remains pegged at the hard limit (as in the previous analysis; Tables \ref{fits} and \ref{full_band}), thus consistent with requiring no relativistic broadening. This therefore means that the emissivity index has no influence on the fit and is unresolved. The lower limit on the inner radius is virtually unchanged by the free emissivity parameter ($\textgreater319$ versus $\textgreater320r_{\rm isco}$; Table \ref{full_band}). Observation 2 prefers an index of 3.5, but is unable to constrain the emissivity within the $2-3.5$ range. Again this is not surprising given the small relativistic broadening required to fit the narrow profile. Despite the emissivity index being free the constraint on the inner radius is again barely affected ($214^{+212}_{-108}\,r_{\rm isco}$). In Observation 3 the broader profile now allows an index of $\textgreater2.7$ to be constrained, whilst the inner radius shifts slightly  ($129^{+16}_{-24}\,r_{\rm isco}$), but is still consistent with the value determined using a fixed index of $3$.

The emissivity index of Observation 4 is found to be $\textless2.5$, thus being the only observation to require an index less than the Euclidean value of 3. This may represent some change in the corona, possibly linked to the imminent state transition. Due to the decreased centralisation of the emission, where the relativistic effects are stronger, the moderately broad profile is recovered by a decrease in the inner radius to $12^{+14}_{-3}\,r_{\rm isco}$, confirming the degeneracy described before. Nevertheless, this still represents a significant level of truncation in this observation and only serves to isolate it further from the more truncated hard state observations at lower luminosities. Therefore, our decision to fix the emissivity index at 3 has little influence on the trend presented in Fig.\,\ref{Fig:rin_lum}.

\begin{table}
\centering
\caption{The source height $h$ of the corona fitted for an inner accretion disc at the ISCO using the model \textsc{relconv-lp}.}
\begin{tabular}{lcccc}\hline\hline
Parameter		& 1				& 2					& 3				& 4						\\\hline	
$h (r_{\rm g})$		& $\textgreater 92$	& $\textgreater 94$		& $\textgreater 98$	& $\textgreater 86$			\\
$\Delta\chi^2$		& $+65$			& $+39$				& $+116$			& $0$					\\\hline
\end{tabular}
\tablefoot{The parameter $\Delta\chi^2$ refers to the additional $\chi^2$ in comparison to fitting a truncated disc model for the same degrees of freedom (Table \ref{full_band}). In all cases the source height $h$ pegged at the hard limit of $100\,r_{\rm g}$, therefore only lower limits were calculated.}
\label{full_lp}
\end{table}

Recently \cite{Fabian14} performed simulations to investigate how the nature of the corona affects the determination of the inner radius, concluding that the profile width is strongly dependent upon the coronal height (see also \citealt{Dauser13}). Ultimately, increasing the coronal height leads to a reduction in the fraction of photons illuminating the inner regions of the disc, and hence a narrow profile could be a signature of coronal elevation rather than disc truncation. We apply the same lamp-post illumination model \textsc{relconv-lp} (replacing \textsc{relconv}) employed in the simulations of \cite{Fabian14} to investigate whether the height of the corona can alone describe the narrow profiles found in all four hard state observations. We fix the inner radius of the disc to be at the ISCO, and allow the height of the corona $h$ to be fit freely between $2-100\,r_{\rm g}$. The photon index is tied to that of the \textsc{powerlaw} model. All observations prefer a source height larger than $85\,r_{\rm g}$, whilst no fit is improved and is significantly poorer for Observations 1-3 (Table \ref{full_lp}). To paraphrase, the source height alone cannot replicate the narrow line profiles observed as well as a truncated disc. Fixing the black hole spin to be zero (\emph{i.e.} a larger ISCO) yielded the same result. As discussed in \cite{Fabian14}, such a large source height is unlikely, and in conjunction with the poorer fit emphasises the presence of inner disc truncation in the hard state of GX 339-4.


\section{Discussion}
\label{discussion}
In this work we have analysed the Fe line region of GX 339-4 and applied it as a probe of how the inner accretion disc evolves, presenting evidence that the Fe line profile changes with luminosity. Our results indicate that even at relatively high luminosities the inner disc radius is still somewhat truncated, although following a trend of decreasing radius with increasing luminosity. This brings into question at which point, if any, the inner disc has reached the ISCO in the canonical hard state. Such a finding has a direct impact upon the use of reflection features to measure BH spin. The entire range of prograde BH spin is spanned within the inner 5 gravitational radii of the disc, and hence even a slight truncation, much smaller than the extent we present, will have a profound impact upon the confidence of spin determination.

We investigate the use of both the line and self-consistent reflection fitting methods, and the disc is found to be significantly more recessed using the latter technique. Self-consistent reflection modelling takes into account Comptonisation and multiple ionisation stages, which themselves broaden the line profile. These effects are a consequence of disc ionisation, and hence a degeneracy could occur between relativistic broadening, for example the inner disc radius, and the ionisation parameter. We also record a trend of increase in the ionisation parameter with luminosity. On the HID (Fig.\,\ref{Fig:HID}) the hard state track is near-vertical, keeping a roughly constant hardness. Most of the power thus is consistently going mainly into the hard, illuminating, component rather than the soft disc. By definition the ionisation parameter is proportional to the flux of the illuminating component for a constant density, and therefore it should be expected to increase with luminosity.

One major issue currently with the publicly available reflection models is that they describe the illumination of an otherwise cold slab of gas. This assumption is acceptable for use in fitting the spectra of AGN where the disc is relatively cold. However, in the case of BHXRBs the hotter surface layers of the disc will have a significant effect upon the reflection spectrum \citep{Ross07}. Additionally, we have the presence of the thermal disc component in the soft bandpass which may be mistaken as reflection by the model. This problem is further compounded by better statistics at lower energies, which can then drive the reflection fit in this range, rather than through the more revealing Fe line region. We remove the latter issues by fitting only above the energy range where the disc is significant, and hence our continuum description is a much simpler absorbed power-law. The effect of a hotter disc is not so straightforward to determine. As discussed in \cite{Ross07}, the increased ionisation has clear broadening effect on the line profile; however, they focused upon the soft state when the disc emission dominates the soft X-ray spectrum. Studies of the hard state using soft X-ray instruments have shown that the disc flux is low at this stage, and the X-ray spectrum is dominated by the power-law component \citep{Reis10,Kolehmainen13}. Therefore, the effect of ionisation due to the disc flux should not differ substantially between the observations in this study.

It is, however, significant, and likely to be treated in two ways by our model. Either the ionisation parameter, or the relativistic effects, will be increased to reproduce the additional broadening in the line profile. If the outcome is the latter, we should yield inner disc radii values closer to the BH than would be apparent for a self-consistently modelled BHXRB. Therefore, in such a case, we expect the disc to be more truncated than what we record in this investigation.

Although we have displayed distinct evolution of the inner disc in the hard state of GX 339-4, we are still currently limited by our sampling of other BH sources. Many sources show similar behaviour (see {\it e.g.} \citealt{Dunn10}); however, some do not show full state transitions ({\it e.g.} Swift J1753.5-0127; \citealt{Soleri13}), and hence it may be the case that the inner disc evolves differently in the hard state of other BHs.


\subsection{How well constrained is the inner radius?}\label{rtest}
Much of this paper has focused on how the inner radius of the disc is evolving in the hard state. If we just examine absolute values, then this would suggest that the disc is always substantially truncated. However, given how degenerate the unknown inclination parameter is with the inner radius (see \S\ref{inclination}), we cannot be sure of this, and instead concentrate more on the trend between observations. It should also be noted that other works have found contrasting results to those found here. For example, \cite{Shidatsu11} analysed three \emph{Suzaku} observations at a very similar luminosity to Observation 2, and found a smaller inner radius (although not significantly) using the line model \textsc{diskline}. However, they also fit a value for $q$ of 2.3, which will force the inner radius to be smaller to broaden the model profile. They also re-analyse Observation 1 and show that the inner radius is larger, although again not as large as what we uncover because of the smaller $q$ that they adopt. In fitting a feature with such a small signal, model degeneracies can easily affect absolute values of Fe line fitting, and the important fact is that \cite{Shidatsu11} find the same trend to us (lower luminosity--larger inner radius).

At first sight, the magnitude of the confidence limits on the inner radius would suggest that our analysis is uncertain, perhaps even flawed, but this is simply due to the decreasing relativistic effect at large radii. For example, the fits to the Fe K$\alpha$ line with \textsc{relline} (Table \ref{fits}) find confidence limits a factor of over 100 larger for Observation 1 than Observation 4. The fit is just as well constrained, and as well as the other parameters, it is just the smaller impact on the spectrum at large radii that drives the increasing limits.

Although we are not focusing upon the absolute values of the inner radius, we are still interested in how well they represent the true value. One test of this is to calculate whether the EW we observe matches that expected for a reflecting slab truncated to the radius we infer. A disc will subtend a solid angle of $\Omega=2\pi(\cos{\theta_{\rm in}}-\cos{\theta_{\rm out}})$, where the respective angles are to the radii corresponding to the inner and outer radii from the centre of the disc. These are calculated as $\theta=\arctan{(R/h)}$, where R is the respective disc radius, and h is the height of the illuminating source above the disc, for which we assume a height of $20\,r_{\rm g}$. We note that this value is uncertain, but is a reasonable estimate. Given that R$_{\rm out}\gg h$ we can use $\theta_{\rm out}=90$. We estimate expected EW values from \cite{Garcia13} using the values for the ionisation parameter we fitted using \textsc{xillver} (Table \ref{fits}). Assuming isotropic emission, the ratio of the EW calculated with \textsc{relline} to the expected value should be equivalent to $\Omega/2\pi$.

We find reasonable agreement in each case, with the EW ratio factors of 1.17, 0.94, 0.82 and 1.14 larger than the solid angle for Observations 1 to 4, respectively. Each observation is consistent with the EW ratio within the limits on the inner radius parameter. Although this is a good indication that our absolute value of the inner radius is reasonable, some uncertainties still remain. For example, the expected EW is calculated from the whole Fe K region, where Fe K$\beta$ emission could contribute up to $\sim15\%$ of the EW depending on the ionisation stage. Also, flaring of the outer disc is expected (see {\it e.g.} \citealt{C-S13} and references therein), which could significantly increase the EW observed.


\subsection{On the effect of the spin parameter}\label{spin_discuss}
Throughout this analysis we have assumed that the spin of the BH in GX 339$-$4 is a=0.9. This is the upper limit found by \cite{Kolehmainen10} using the continuum method. It is also close to values obtained previously through reflection fitting \citep{Reis08,Miller08}. Since the spin parameter is inferred from the ISCO, fitting the inner radius and spin parameters can potentially be highly degenerate. Therefore, since the aim of this investigation was to measure the inner disc radius, we fixed the value of the spin to be a=0.9 to keep all the analysis consistent.

\cite{Yamada09} suggested that the spin of GX 339$-$4 may be smaller than 0.9, but defining a spin value has, in fact, very little effect upon the conclusions of this work. For a fixed radius, increased spin will have some effect due to frame dragging, but this is quite minimal, and only observable for the very inner regions of the accretion disc \citep{Dauser10}. Since the disc is found to be largely truncated, the adopted spin value is expected to have little or no effect upon the results of this study.

\section{Conclusions}
\label{conclusions}
We have systematically analysed how the inner accretion disc evolves in the canonical hard state of GX 339-4 using the Fe line region. Our results have shown that the inner accretion disc moves closer to the black hole at higher luminosities, consistent with the broader profile found in the spectra. The poor fit to each spectrum using a single line model points to a reflection spectrum not dominated by relativistic effects, with significant broadening also arising as a result of ionisation. When this is taken into account we find an improved fit and significantly larger inner radii, concluding that the inner disc is truncated throughout the hard state.

The trend we find thus fully supports the truncated disc model of the hard state. Furthermore, when extending our analysis to the hard-intermediate state we find even smaller inner radii confirming that the inner disc is truncated throughout the entire hard state. It has also been suggested that the break frequency found in the power density spectra of the hard state corresponds to the inner radius of the disc \citep{Gilfanov99}. Upon comparing this to our results from spectral fitting we find a remarkably similar trend. Together, these two independent conclusions provide very strong evidence in favour of disc truncation in the hard state. Additionally, whilst it has been suggested in \cite{Fabian14} that the coronal height can reproduce the profile of a truncated disc, we show that the observations considered in this study unanimously require that the inner accretion disc is truncated. 

This result implies that the current sample of spin estimates in the hard state are inaccurate. Therefore, any distinct conclusions drawn from these estimates, such as the spin-powering of relativistic jets, may well be biased.


\section*{Acknowledgements}
The authors are very grateful to Javier Garcia for the use of the reflection model \textsc{xillver}, and Stefano Bianchi for useful discussion. We also thank the referee for careful and constructive comments that have undoubtedly improved this article. This research has made use of the General High-energy Aperiodic Timing Software (GHATS) package developed by T.M. Belloni at INAF - Observatorio Astronomico di Brera. It has also made use of data obtained with the \textit{XMM-Newton}, \textit{Suzaku} and \textit{RXTE} satellites. DSP acknowledges financial support from the STFC. GP acknowledges support via an EU Marie Curie Intra-European fellowship under contract no. FP-PEOPLE-2012-IEF-331095. TMD acknowledges funding via an EU Marie Curie Intra-European Fellowship under contract no. 2011-301355.

\bibliographystyle{aa}
\bibliography{refs}

\end{document}